\documentclass[iop,revtex4]{emulateapj}

\usepackage{amsmath}

\newcommand{\mytilde}{\raise.19ex\hbox{$\scriptstyle\sim$}}
\newcommand{\symvec}[1]{\mbox{\boldmath${#1}$}}

\shorttitle{DLS Cosmic Shear Tomography}
\shortauthors{Jee et al.}

\begin{document}

\title{COSMIC SHEAR RESULTS FROM THE DEEP LENS SURVEY - II: FULL COSMOLOGICAL PARAMETER CONSTRAINTS FROM TOMOGRAPHY} 

\author{M. JAMES JEE\altaffilmark{1,2}, J. ANTHONY TYSON\altaffilmark{2}, STEFAN HILBERT\altaffilmark{3,4}, MICHAEL D. SCHNEIDER\altaffilmark{5}, \\ SAMUEL SCHMIDT\altaffilmark{2}, AND DAVID WITTMAN\altaffilmark{2}}

\begin{abstract}
We present a tomographic cosmic shear study from the Deep Lens Survey (DLS), which,
providing a limiting magnitude $r_{\mathrm{lim}}\sim27$ ($5\sigma$),
is designed as a pre-cursor Large Synoptic Survey Telescope (LSST) survey with an emphasis on depth.
Using five tomographic redshift bins, we study their auto- and
cross-correlations to constrain cosmological parameters. We use a luminosity-dependent nonlinear 
model to account for the astrophysical systematics originating from intrinsic alignments of galaxy shapes.
We find that the cosmological leverage of the DLS is among the highest among existing $>10$ sq. deg cosmic shear surveys. Combining the DLS tomography with the 9-year results of the Wilkinson Microwave Anisotropy Probe (WMAP9) gives $\Omega_m=0.293_{-0.014}^{+0.012}$, $\sigma_8=0.833_{-0.018}^{+0.011}$, $H_0=68.6_{-1.2}^{+1.4}~\mbox{km~s}^{-1}~\mbox{Mpc}^{-1}$, and
$\Omega_b=0.0475\pm0.0012$ for $\Lambda$CDM, reducing the uncertainties of the WMAP9-only constraints by $\mytilde50$\%.
When we do not assume flatness for $\Lambda$CDM, we obtain the curvature constraint $\Omega_k=-0.010_{-0.015}^{+0.013}$ from the DLS+WMAP9 combination, which however is not well constrained when WMAP9 is used alone.
The dark energy equation of state parameter $w$ is tightly constrained when Baryonic Acoustic Oscillation (BAO) data are added, yielding $w=-1.02_{-0.09}^{+0.10}$ with the DLS+WMAP9+BAO joint probe. The addition of supernova constraints further tightens the parameter to $w=-1.03\pm0.03$.
Our joint constraints are fully consistent with the final Planck results and also the predictions of a $\Lambda$CDM universe.
\end{abstract}

\altaffiltext{1}{Department of Astronomy and Center for Galaxy Evolution Research, Yonsei University, 50 Yonsei-ro, Seoul 03722, Korea}
\altaffiltext{2}{Department of Physics, University of California, Davis, One Shields Avenue, Davis, CA 95616}
\altaffiltext{3}{Excellence Cluster Universe, Boltzmannstr. 2, 85748 Garching, Germany}
\altaffiltext{4}{Ludwig-Maximilians-Universit{\"a}t, Universit{\"a}ts-Sternwarte, Scheinerstr. 1, 81679 M{\"u}nchen, Germany}
\altaffiltext{5}{Lawrence Livermore National Laboratory, P.O. Box 808 L-210, Livermore, CA 94551}

\section{INTRODUCTION \label{section_introduction}}
Subtle changes in distant galaxy images occur as the light bundles traveling to us are deflected by gravitational lensing effects due to intervening large scale structures. 
Precision measurement of these subtle changes, often called ``cosmic shear", provides a powerful method to probe the statistical properties of the large scale structures, which are in turn the results 
determined by
the initial conditions of the universe. Cosmic shear is sensitive to both the geometry of the universe and the growth of its structure, making it one of the most efficient tools for constraining dark energy (Albrecht et al. 2006).

As both cosmic geometry and structure evolve with time, we want to measure the cosmic shear statistics at multiple redshifts. This tomographic analysis becomes possible if a survey can access high-redshift galaxies with low surface brightness limits. Therefore, when designing a cosmic shear survey, given a fixed amount of telescope time, one faces a critical decision regarding the balance between area and depth. 

The Deep Lens Survey (DLS)\footnote{http://dls.physics.ucdavis.edu} is a 20 sq. deg cosmic shear survey with an emphasis on depth reaching $\mytilde27^{th}$ mag in $R$ at the 5$\sigma$ level, and uniformly good image quality.
The images were taken over $\mytilde120$ nights on NOAO's Blanco and Mayall 4-meter telescopes. If the same telescope time had been used to cover a 150 sq. deg instead, the corresponding limiting magnitude would have been $\mytilde24.5^{th}$ mag, which is similar to the expected final depth of the on-going Dark Energy Survey (DES)\footnote{
http://www.darkenergysurvey.org/}
or the mean depth of the Canada-France-Hawaii-Telescope Lensing Survey (CFHTLens; Erben et al. 2013). 

This deliberate choice of ``depth over area" was made after considering the following five issues. First, it enables us to obtain a longer redshift baseline to break parameter 
degeneracies. For example, shallower surveys with wider area are not efficient in shrinking the length of the ``banana" in the $\Omega_m-\sigma_8$ plane because of the $\sigma_8 \Omega_m^a=\mbox{constant}$ degeneracy.
Second, it compensates for the loss of volume due to a reduced area by extending the volume along the line of sight. An important source of statistical uncertainty in the existing moderate-scale ($<200$ sq. deg) cosmic shear surveys is the sample variance due to a limited volume. A common misconception is that a wide survey contains a much higher volume than a deep survey with a smaller area, but this ignores the volume extension along the line of sight in a deep survey. As an illustration, we can consider
two surveys with a mean source redshift of z=0.7 and 1.1. The difference in co-moving volume (bounded by the mean source redshift) is nearly a factor of 3.
Third, it boosts the amplitude of the cosmic shear signals due to the varying strength of the lensing kernel. Given the same lens mass, the strength of lensing signals is a simple function of the two distances: observer-to-lens versus lens-to-source. For example, when a lens is at $z=0.5$, the shear at $z=1.1$ is nearly twice that at $z=0.7$, leading to a factor of 4 difference in correlation function amplitudes. Thus, a deep survey can take advantage of the geometric effect of gravitational lensing more efficiently.
Fourth, deep surveys provide more galaxies per area. We should clarify that this merit does not entirely overlap with the first point. In addition to new sources at higher redshift, the increased limiting magnitude allows us to detect fainter galaxies at a given redshift. Needless to say, the increased source density reduces shape noise caused by intrinsic ellipticity dispersion.
Fifth, it mitigates the effects of intrinsic alignments. Intrinsic alignments of galaxies can be an important theoretical systematic in precision cosmic shear. Current studies \citep[e.g.][]{2013MNRAS.432.2433H} indicate that their effects are dominated by luminous red galaxies (LRGs). Because a deep survey can access fainter galaxies, the fraction of the LRGs decreases as we increase the limiting magnitude. Following the approach of \citet{2011A&A...527A..26J}, we estimate that at $z=0.5$ the amplitude of the IA power spectrum is reduced by a factor of two as we increase the limiting magnitude from $r_{\mathrm{lim}}\sim24$ to $\mytilde27$. These considerations drove the cadence and survey details of the DLS. 

Of course, there are a few potential downsides in deep imaging cosmic shear surveys. If one chooses a small contiguous field, the increase of noise from the cosmic variance
can offset the merits mentioned above. In DLS, we tried to reduce the cosmic variance by choosing the five 4 sq. deg fields widely separated.
Another issue is object blending. With ground-based seeing, going fainter than $\mytilde28$ mag results in a non-negligible increase in the number of projected overlapping galaxies. We designed the DLS to reach ~27 mag at the five sigma level to mitigate the issue.


Galaxy shear estimation and control of PSF systematics is crucial. The methods developed for the DLS are discussed below. The natural first step in a cosmic shear analysis of survey data is a simple two-dimensional (2D), non-tomographic study, including various null tests for shear systematics. In Jee et al. (2013) [hereafter J13], we presented our 2D analysis of the DLS cosmic shear. That study shows that the shear systematics are controlled to well below the statistical uncertainties based on residual point-spread-function (PSF) correlations, B-mode statistics, and star-galaxy correlations.
We also demonstrated that, when combined with Cosmic Microwave Background (CMB) data, the DLS cosmic shear can put tight constraints on the two parameters: matter density $\Omega_m$ and normalization $\sigma_8$. Combined with the 7-year result of the Wilkinson Microwave Anisotropy Probe (WMAP7; Komatsu et al. 2010), the DLS 2D study provides $\Omega_M=0.278\pm0.018$ and $\sigma_8=0.815\pm0.020$, which reduces the WMAP7-only uncertainties by $\mytilde31$\% and $\mytilde29$\%, respectively.

In this paper, we present full cosmological parameter constraints from the DLS tomography, combining three other cosmological probes. Specifically, we choose the 9-year result of the Wilkinson Microwave Anisotropy Probe (WMAP9; Hinshaw et al. 2013) for CMB data. Although selecting the Planck CMB results (Planck Collaboration et al. 2015; hereafter Planck2015-CMB) instead of WMAP9 leads to smaller parameter uncertainties, we make the above decision because using WMAP9 as an external data set provides an interesting opportunity to compare the final results against the Planck2015-CMB results, which by themselves produce considerably tight constraints and are reported to be in tension with some cosmology studies (e.g., MacCrann et al. 2015).
For supernova data, we utilize the publicly available union2.1 compilation provided by Suzuki et al. (2012). For Baryonic Acoustic Oscillation data, we select the results from the following four studies: Beutler et al. (2011), Padmanabhan et al. (2012), Anderson et al. (2012), and Blake et al. (2012).

Joint probes are critical in constraining cosmological parameters. First of all, they enable the diagnosis and mitigation of systematics. Because each cosmological probe is subject to different bias arising from both theoretical and instrumental systematics, combining multiple probes can turn these systematic errors into statistical errors.
Second, joint probes are required to break parameter degeneracies. For example, the degeneracy direction of the $\Omega_m$ versus $\sigma_8$ combination in cosmic shear studies is nearly perpendicular to that in CMB studies.
Third, some cosmological parameters are inaccessible in one type of probe. For example, the $\sigma_8$ parameter is not constrained in probes using purely geometric effects such as supernovae and BAOs. The dark energy equation of state parameter $w$ affects both geometry and growth, but it has a minimal effect on the features in the early universe (at the last scattering surface) probed by CMB missions.

We present our tomographic cosmic shear results as follows. In \textsection\ref{section_dls} we describe the DLS data and the shear estimation algorithm. A brief review on theoretical formalism regarding cosmic shear tomography and intrinsic alignments is given in \textsection\ref{section_theory}. 
\textsection\ref{section_analysis} describes the tomography setup and discusses the basic auto- and cross-correlation functions. We present full cosmological parameter constraints in \textsection\ref{section_parameters}.
The impact of intrinsic alignments on parameter constraints and comparisons with other cosmological probes
are discussed in \textsection\ref{section_discussion} before we conclude in \textsection\ref{section_conclusion}.

\section{THE DEEP LENS SURVEY}\label{section_dls}

\subsection{The Deep Lens Survey Data}
The DLS is designed as a pre-cursor Large Synoptic Survey Telescope (LSST)\footnote{http://www.lsst.org} survey with an emphasis on depth reaching a 5$\sigma$ limiting magnitude of $\mytilde27^{th}$ mag in $R$.
The survey covers five 2$\degr\times2\degr$ fields (hereafter F1-F5).
Two fields (F1 and F2) are in the northern sky and were observed with the Kitt Peak Mayall 4 m telescope/Mosaic Prime-Focus Imager (Muller et al. 1998).
The rest (F3, F4, and F5) are in the southern sky, using the Cerro Tololo Blanco 4 m telescope/Mosaic Prime-Focus Imager.
The DLS data consist of a total of 140 nights of {\it B, V, R,} and {\it z} imaging. Priority was given to the {\it R} filter, where we measure our lensing signal, whenever the seeing was better than $\mytilde0.9\arcsec$. Thus, the survey depth and resolution were designed to minimize point spread function (PSF) systematics.
The $R$-band survey depth (limiting magnitude) was set based on our image simulation result, which shows that the removal of the PSF effect in shape measurement through forward-modeling improves at faint limiting magnitude and remains stable down to $R\approx26^{th}$.
The mean cumulative exposure time in $R$
is $\mytilde18,000$ s per field whereas it is about 12,000 s per field for each of the rest of the filters. 
The approximate resulting limiting magnitudes (at the $5\sigma$ level) are 26, 26, 27, 26 in {\it B, V, R,} and {\it z}.  Our conservative selection scheme provides reliable galaxy shapes down to $R\sim26$ with a mean redshift of $z\sim1.2$. 
The resulting
raw number density is $n\sim17$ per sq. arcmin.
When galaxies are weighted according to their inverse noise variance (combining both shape dispersion and measurement noise),
the effective number density of the source population becomes $n_{\mathrm{eff}}=11$ per sq. arcmin (Equation 7 of Jee et al. 2014).

\subsection{Shear Estimation}
Conventional galaxy shear estimation based purely on galaxy shape and PSF shape measurements on co-added image stacks incurs dangerously large PSF systematics and associated B-mode. 
This is because in general the PSF varies smoothly on the individual exposures but not on the stack.
Therefore, a high-fidelity PSF map must first be obtained from the individual exposures that go into the deep stack, and then the final PSF variation 
on the deep stack must be modelled after carefully taking into account  rotations/shifts, flux calibrations, flat-fielding, masks, etc.

We used the principal component analysis method of Jee \& Tyson (2011) to model the PSF and its variation on individual frames. 
We stack these PSF models at the exposure level and obtain the final high resolution PSF models for our co-adds.
In J13, we showed that our residual PSF correlation amplitudes are negligible, based on the statistics suggested by Rowe (2010).

Accurate PSF modeling is only a necessary condition for precision shear estimation. There are a number of subtleties in shear estimation that bias the results. For example, the mismatch between analytic and true
profiles in the galaxy surface brightness distribution can certainly lead to non-negligible bias. This
so-called ``model-bias" is expected to increase for deeper surveys, where more high-$z$ (thus more irregular) galaxies comprise the source population. Also, a nonlinear mapping from pixel noise to parameter uncertainties has been shown to be critical (e.g., Refregier et al. 2012). This so-called noise-bias can lead to shear bias as high as $\mytilde10\%$.
Imperfect de-blending is receiving growing attention in deep surveys, where a significant fraction of sources overlap with neighboring objects. Recently, based on image simulations, Hoekstra et al. (2015) found that even in relatively shallow surveys this de-blending bias is still non-negligible. We agree that their finding is reasonable because the survey depth does not alter the flux ratio between two neighboring objects although the fainter of the two can be missed in source detection procedure in shallow surveys. Apart from the aforementioned sources, while participating in GREAT3, we find that many other details such as the size of the postage stamp images (we refer to this as ``truncation bias" below), handling of the central pixel, initial size and ellipticity of the PSF, etc., will matter in future cosmic shear surveys, whose requirements in shear estimation precision is well below $\mytilde0.1$\%\footnote{In terms of the so-called multiplicative $\gamma_m$ and additive $\gamma_c$ bias parameters, Massey et al. (2013) estimate $\gamma_m=2\times10^{-3}$ and $\gamma_c=2\times10^{-4}$}.

Our principle to address the above issues is statistical shear calibration through high-fidelity image simulations. This means that we 
measure shear bias by comparing recovered shears with true values through the procedures closely mimicking the real image analysis. The primary merit of the approach is that many subtleties are automatically included and thus cancelled. For example, we do not have to explicitly single out the truncation bias as long as both simulation and real data analysis use the same postage stamp size. 

In DLS, we measure galaxy shapes by fitting an elliptical Gaussian. One can argue that elliptical Gaussian may not be an optimal choice for real galaxies, creating rather large model-bias compared to more sophisticated models such as Sersic or double-sersic models. However, sophisticated models 
give large uncertainties for shape parameters due to many other nuisance parameters, which in turn increase noise bias. Therefore, currently it is unclear that highly sophisticated galaxy models are indeed superior. Perhaps, one can consider a single Sersic model as a reasonable compromise.
In fact, we participated the GREAT3 challenge with the method implementing the Sersic galaxy fitting, and the performance turned out to be excellent.
Thus, in future studies, it will be a useful exercise to experiment with the method for analyzing DLS data, although the task will involves new suites of time-consuming image simulations. In this paper, we utilize our previous efforts in J13.

We perform our shear calibration based on our LSST image simulation tools (Jee \& Tyson 2011), where the Ultra Deep Field (UDF; Beckwith, et al. 2006) galaxies are used. 
Although one can separate the aforementioned issues (model-bias, noise-bias, etc.), we choose to investigate only their combined effect and characterize it using a single parameter $R$-band magnitude:
\begin{equation}
m_{\gamma}=6\times10^{-4} (m_R - 20 )^{3.26} + 1.04. \label{eqn_m_gamma}
\end{equation}
\noindent
Because we conserve the statistical properties of the galaxies as a function of magnitude in UDF (i.e., we do not ``clone" faint galaxy images using bright high S/N galaxy images), various sources of bias should be taken care of in this step. For example,
both model- and noise-bias should be characterized by the apparent magnitude proxy. Also, the effect of decreasing galaxy half light radius with $z$-bin relative to the PSF is statistically captured.

Nevertheless, in future studies, it would be useful to test a calibration model with explicit $z$-bin variables including
morphology, size, and noise.
In our case, the calibration parameter $m_{\gamma}$ is always greater than unity and should be multiplied by our raw ellipticity.
This equation shows that the mean shear is biased low by $\mytilde4$\% at $m_R=22$, and the bias can rise as high as $\mytilde24$\% at the faint end ($m_R=26$). 
In our 2D cosmic shear analysis (J13), we demonstrated that based on star-galaxy correlations and ``B-mode" analysis, the residual shear systematics are negligible.
The fidelity of the DLS shape measurement scheme has been tested in the third GRavitational lEnsing Accuracy Testing (GREAT3), and our method (called SFIT) won the challenge (Mandelbaum et al. 2015).

Using eqn~\ref{eqn_m_gamma}, we derive the mean value of $m_{\gamma}$ separately for each tomographic bin and apply the calibration to shear correlation functions. As was done in J13, we assume that our shear calibration is uncertain at the 3\% level and thus marginalize over this interval in our cosmological parameter estimation. Similar to our handling of photo-z bias, the residual shear bias correlation between different tomographic bins is assumed to be 100\%, which is a conservative choice because this perfect positive correlation assumption will maximally degrade the cosmological parameter uncertainties.

\subsection{Photo-$z$ Estimation}

In \textsection\ref{section_analysis} we will describe breaking the galaxy data into tomographic bins using photometric redshifts (photo-$z$'s). For DLS, photo-$z$'s are calculated using BPZ (Benitez 2000). Schmidt \& Thorman (2013) carefully calibrate the priors and ``tweak" the spectral energy distribution (SED) templates utilizing $\mytilde10,000$ spectroscopic redshifts from the HEctospec Lensing Survey (SHELS; Geller et al. 2005).  However, it is crucial to verify that the utility of this customization based on the specific sample does not result in overtraining.
Schmidt \& Thorman (2013) and J13 compared the DLS photometric redshifts with the spectroscopic redshifts from the PRIsm MUlti-object Survey (PRIMUS; Coil et al. 2011), an independent survey partially overlapping DLS, and demonstrated that the agreement is excellent.

In a galaxy-galaxy lensing study from the DLS, Choi et al. (2012) further examined the DLS photo-$z$ systematics by swapping foreground and background photo-z bins as a null test, and found that the results were consistent with predictions from known systematics. They also calculated angular cross-correlations $w(\theta)$ between different photo-$z$ bins and concluded that the ratios between the auto- and cross-correlations of these samples were less than 0.1, consistent with the overlap of the $p(z)$ tails and the known photo-$z$ outlier rates.

In J13, we demonstrated that the photo-z bias is small ($\mytilde1$\%) in the low redshift regime ($z\lesssim1.2$), where many spectroscopic redshifts are available. It is extremely difficult to estimate the bias in the high ($z \gtrsim1.2$) redshift regime where not many spectroscopic redshifts within DLS exist. The role of the used prior might be significant here, as objects with poor SED constraints are likely to default to the prior. In J13, we estimated the prior dependence by separately running BPZ using the Hubble Deep Field North (HDFN) prior and comparing the results with the ones obtained with the VIMOS-VLT Deep Survey (VVDS; Le F{\`e}vre et
al. 2005) prior, which is the main prior employed in DLS. The resulting ~3\% change in the mean redshift was adopted as the photo-z bias in J13, and thus the cosmological parameters were obtained after marginalization over the 3\% interval.
We believe that this estimate is conservative because the HDFN prior is derived from a single {\it Hubble Space Telescope (HST)} field ($\mytilde2\arcmin\times2\arcmin$), whereas the main VVDS prior is obtained from a much larger (by a factor of $\mytilde2000$) survey. As was done in J13, in the current tomographic analysis, we also marginalized over the same 3\% bias interval (flat prior) in photometric redshift. We assumed that the photo-z bias parameters in the five bins are 100\% correlated, which gives the maximum deviation and is certainly more conservative than the assumption that the correlation is weaker or independent. We refer readers to J13, Schmidt \& Thorman (2012), Wittman et al. (2012), and Choi et al. (2012) for further details on photometric calibration, weak-lensing shape measurements, and photometric redshift estimation.

\section{THEORETICAL BACKGROUND} \label{section_theory}
   \subsection{Cosmic Shear Tomography}
Cosmic shear measures a subtle correlated distortion of galaxy shapes due to gravitational lensing by the large scale structures of the universe. Many statistics have been suggested to quantify the correlation and draw cosmological information out of the statistics. In this paper we use two-point correlation statistics to constrain cosmological parameters.

The two-point correlation between galaxies $i$ and $j$ at a separation $\theta$ is defined as:
\begin{equation}
\xi_{tt} (\theta) = \frac{\Sigma_{i,j} w_i w_j e_{t,i} e_{t,j} } {\Sigma w_i w_j} \label{eqn_xi_tt}
\end{equation}
and
\begin{equation}
\xi_{\times\times} (\theta) = \frac{\Sigma_{i,j} w_i w_j e_{\times,i} e_{\times,j}} {\Sigma w_i w_j}. \label{eqn_xi_xx}
\end{equation}
\noindent
where the summation is carried out over every possible combination of $i^{th}$ and $j^{th}$ galaxies ($i< j$).
The subscript $t$ refers to the projection of the galaxy ellipticity perpendicular to the line connecting
the galaxy pair. The same is true for the subscript $\times$ except that the projection is along the 45$\degr$ angle
with respect to the connecting line.
The weight associated with the ellipticity of the $i(j)^{th}$ galaxy is the inverse shape noise variance $w_{i (j)}$, which is defined as
\begin{equation}
w_i  = \frac {1} { \sigma_{SN}^2 + (\delta e_i)^2} \label{eqn_shear_weight},
\end{equation}
\noindent
where $\sigma_{SN}$ and $\delta e_i$ are the ellipticity dispersion and
measurement error, respectively.

In this paper we use the following two linear combinations of $\xi_{tt}$ and $\xi_{\times\times}$ when comparing to prediction:
\begin{equation}
\xi_+ = \xi_{tt} + \xi_{\times\times}
\end{equation}
and
\begin{equation}
\xi_- = \xi_{tt} - \xi_{\times\times}.
\end{equation}

The predicted values of these two functions $\xi_+$ and $\xi_-$ are easily evaluated by convolving the convergence mass power spectrum $P_\kappa$
with the zeroth $J_0$ and fourth $J_4$ order Bessel functions of the first kind, respectively (Bartelmann \& Schneider 2001). That is,
\begin{equation}
\xi_{+,-}^{k l} (\theta) = \frac{1}{2 \pi} \int_0^{\infty} d \ell \ell J_{0,4} (\ell \theta) P_{\kappa}^{k l} (\ell) \label{eqn_xi},
\end{equation}
\noindent
 where $\ell$ is an angular wave number. The superscripts $k$ ($l$) refers to the $k^{th}$ ($l^{th}$) redshift shell.
  The convergence mass power spectrum $P_\kappa^{kl}$ is
 \begin{multline}
 P_\kappa^{kl} (\ell) = \frac{9}{4} \Omega_M^2 \left ( \frac{H_0}{c} \right )^4 \int_0^{\chi_{\mathrm{max}}} d \chi 
\\\times
\frac{ g_k (\chi) g_l (\chi) }{a^2 ( \chi ) } P_{\delta}  \left ( \frac{\ell}{f_K (\chi)}, \chi \right ),
\end{multline}
where $H_0$ is the Hubble parameter, $\Omega_M$ is today's matter density, $a (\chi) $ is the scale factor at a redshift corresponding to a comoving distance $\chi$, $f_K$ is the
comoving angular diameter distance, $P_{\delta}$ is the matter power spectrum,
and $g_k$ is the lensing efficiency factor:
\begin{equation}
g_k(\chi) = \int_{\chi}^{\chi_{\mathrm{max}}} d \chi^{\prime} p_k (\chi^{\prime}) \frac{ f_K (\chi^{\prime} - \chi) }{f_K (\chi^{\prime} ) }
\end{equation}
for the $k^{th}$ redshift shell with a redshift distribution $p_k(\chi)$. We remind readers that in our non-tomographic study (J13) we used only a single broad redshift shell.
 
\subsection{Intrinsic Alignment Model} \label{section_IA_model}
Conventionally one of the fundamental posits of weak-lensing is zero or sufficiently small correlation of  galaxy shapes in the absence of gravitational 
lensing; i.e. intrinsically random galaxy ellipticities.
However, both theory and observation have shown that this assumption is not valid
at the level of precision that we soon will achieve. This intrinsic alignment, unless carefully addressed, may become a critical factor that limits our interpretation of cosmic shear data from the next generation surveys such as LSST, Euclid\footnote{http://www.euclid-ec.org}, WFIRST-AFTA\footnote{http://wfirst.gsfc.nasa.gov}, etc.

A straightforward shape correlation occurs between physically close pairs experiencing the common tidal force from the same environment. 
This correlation, often termed ``II" (intrinsic-intrinsic), is positive on average and hence artificially increases amplitudes of shear-shear correlations (Catelan et al. 2001). 

A less straightforward, but greater systematic error in cosmic shear from intrinsic alignment is a correlation between gravitational shear and intrinsic ellipticity, which is often referred to as ``GI" (gravitational-intrinsic). This correlation arises because images of background galaxies are sheared by  gravitational lensing effects of foreground large scale structures, inside which
the galaxies are intrinsically aligned, but are physically at great distances from the background galaxies along the line of sight (Hirata \& Seljak 2004). Since in general  lensing efficiency is greater for larger separation between lens and source along the line of sight, this GI systematic is redshift-dependent and becomes substantial when  we cross-correlate galaxy shapes with a large difference in redshift.

Both II and GI effects can in principle become more important in tomographic studies than in two-dimensional ones. In tomographic cosmic shear, there is a  higher chance that a chosen galaxy pair is physically close when we evaluate auto-correlation functions because the redshift ranges are narrower. Also, in a tomographic study, GI signals become amplified when we cross-correlate between low-z and high-z bins. In two-dimensional studies, these GI and II pairs  are greatly outnumbered by other pairs.

Observational studies show that intrinsic alignments are type-dependent, preferentially stronger in luminous red galaxies (LRGs) as shown by Hirata et al. (2007) and Joachimi et al. (2011). The intrinsic alignment signal from blue galaxies is consistent with zero according to Mandelbaum et al. (2011), who investigate the effect using blue galaxies from the Wiggle-Z survey\footnote{http://wigglez.swin.edu.au/site/}. This significant type-dependence may be utilized to our advantage if we can identify (utilizing photometric and morphological data) and remove LRGs from our source population. Because LRGs comprise a small fraction, the resulting loss in S/N is not substantial in general.

Some have suggested methods that null IA effects. For example, Joachimi \& Schneider (2008) proposed a method to down-weight IA systematics utilizing the aforementioned redshift dependence of the GI signals. However, given the expected photometric redshift uncertainties of near-term surveys, the method likely suffers from a non-negligible loss in S/N, compromising the parameter constraining power of the cosmic shear surveys.

Another solution to address IA systematics is to model the amplitude of the systematic and remove it from observed signals. Although there  exists no theoretically solid model that enjoys the consensus of the community,  the 
merit of this approach is its ability to maximize the S/N of a given survey.
In this paper, we use the luminosity-dependent non-linear intrinsic alignment model of Joachimi et al. (2011) to address the IA systematics in DLS. Below we describe the formalism.

The nonlinear IA model of Joachimi et al. (2011) is based on the linear alignment model of Catelan et al. (2001) and Hirata \& Seljak (2004), which relates the IA power spectrum $P_{\delta I}$ to the matter power spectrum $P_\delta$ via:
\begin{equation}
P_{\delta I} = - A ~C_1~ \rho_c \frac{\Omega_m}{D(z)} P_\delta, \label{eqn_IA}
\end{equation}
\noindent
where $\rho_c$ is the critical density of the universe today, $\Omega_m$ is the matter density, and $D(z)$ is the growth factor (normalized to unity at $z=0$).
$C_1$ and $A$ are coefficients to be determined. $C_1$ is often chosen to be fixed at $C_1=5\times10^{-14} h^{-2} M_{\sun}^{-1} \mbox{Mpc}^3$ motivated by the comparison of the SuperCOSMOS observations with the linear alignment model (Bridle \& King 2007) whereas $A$ is considered to be a free parameter.

The matter power spectrum $P_\delta$ in Equation~\ref{eqn_IA} is a linear power spectrum in the original derivation. However, Bridle \& King (2007) replaced it with a full nonlinear power spectrum. This extension lacks a solid justification. Nevertheless, this nonlinear version has been shown to provide a reasonable match with observations. Heymans et al. (2013) used this nonlinear IA model in their tomographic analysis of the CFHTLenS data while marginalizing over the parameter $A$. Their result with the full sample is consistent with a zero IA signal ($A=-0.48_{-0.87}^{+0.75}$). 

One of the outstanding weaknesses of the above model is a lack of explicit dependence on source galaxy redshift. However, the properties of the source galaxy population, such as the fraction of LRGs, and the strength of large-scale tidal fields causing IAs are redshift-dependent.
Joachimi et al. (2011) investigated these redshift dependencies by combining the MegaZ-LRG (Collister et al. 2007) and SDSS-LRG (Eisenstein et al. 2001) samples.
Following their work, we relate the GI power spectrum $P_{GI}$ to $P_{\delta}$ via:
\begin{equation}
P_{GI}= - A ~C_1~ \rho_c \frac{\Omega_m}{D(z)} P_\delta \left ( \frac{1+z}{1+z_0} \right )^{\eta} \left ( \frac{L}{L_0} \right ) ^{\beta}, 
\end{equation} 
\noindent
where $z_0$ is an arbitrary pivot redshift, and $L_0$ is the pivot luminosity defined as an absolute $r$-band magnitude of $-22$ when passively evolved to $z=0$.
Their best-fit results are consistent with no redshift evolution $\eta=-0.27_{-0.79}^{+0.80}$, but with a strong dependence on source
luminosity $\beta=1.13_{-0.20}^{+0.25}$; Joachimi et al. (2011) presented their best-fit results for five different combinations of samples. We quote the results from the full combination (MegaZ-LRG + SDSS LRG + L4 + L3).  The best-fit amplitude is $A=5.76_{-0.62}^{+0.60}$.

If we define the following:
\begin{equation}
f(z,L)= - A\, C_1 \, \rho_c \frac{\Omega_m}{D(z)} \left ( \frac{1+z}{1+z_0} \right )^{\eta} \left ( \frac{L}{L_0} \right ) ^{\beta}, \label{eqn_fzl}
\end{equation}
\noindent
the power spectrum for the GI and II are expressed as $P_{GI}= f(z,L) P_{\delta}$ and $P_{II}=f(z,L)^2 P_{\delta}$, respectively.
Given a luminosity function and its evolution, we can remove the explicit $L$-dependence from $f(z,L)$. Joachimi et al. (2011) derived
an analytic expression for a survey with a limiting $r$-band magnitude $r_{\mathrm{lim}}$:
\begin{equation}
\frac{ \left < L^{\beta} \right > (z,r_{\mathrm{lim}})} {L_0^{\beta} (z) }  = \left ( \frac{L^*(z)} {L_0 (z)} \right )^{\beta} 
      \frac{\Gamma \left(\alpha+\beta+1, \frac{L_{\mathrm{min}} }{L^* (z)}  \right) } { \Gamma \left(\alpha+1, \frac{L_{\mathrm{min}}}{L^* (z)}  \right)  },
\end{equation}
where $L_{\mathrm{min}}=L_{\mathrm{min}} (z,r_{\mathrm{\mathrm{lim}}})$ is the minimum luminosity of a galaxy at redshift  $z$ for $r_{\mathrm{lim}}$, $L_0(z)$ is the evolution-corrected rest-frame luminosity at $z$ corresponding to an absolute magnitude -22, $\Gamma$ is the incomplete Gamma function, and $\alpha$ is the slope of the luminosity function.
We adopt the result of Faber et al. (2007) for $\alpha$ and the estimation of $L_0(z)$ while ignoring the difference in 
filter throughput between the $r$ and DLS $R$ filters. We refer readers to Appendix C of Joachimi et al. (2011) for other details.

Once we obtain $P_{GI}$ and $P_{II}$, the estimation of the GI and II contribution to $\xi_{+,-}$ becomes straightforward. The projected II and GI power spectra are:
\begin{multline}
 C_{II}^{kl} (\ell) = \frac{9}{4} \Omega_M^2 \left ( \frac{H_0}{c} \right )^4 \int_0^{\chi_{\mathrm{max}}} d \chi 
\\\times
\frac{ n_k (\chi) n_l (\chi) }{a^2 ( \chi ) } 
 P_{II}  \left ( \frac{\ell}{f_K (\chi)}, \chi \right )
 \label{eqn_c_II}
\end{multline}
and
\begin{multline}
 C_{GI}^{kl} (\ell) = \frac{9}{4} \Omega_M^2 \left ( \frac{H_0}{c} \right )^4 \int_0^{\chi_{\mathrm{max}}} d \chi 
\\\times
\frac{ n_k (\chi) g_l (\chi) + n_l (\chi) g_k (\chi)    }{a^2 ( \chi ) } 
 P_{GI}  \left ( \frac{\ell}{f_K (\chi)}, \chi \right ),
\label{eqn_c_GI}
\end{multline}
respectively.
 Note that in Equations~\ref{eqn_c_II} and~\ref{eqn_c_GI} $n_k(\chi)$ is an effective number of sources. Then, the II and GI correlation functions (denoted as 
 $\xi_{\pm}^{kl} (\theta)_{II,GI}$) are:
 \begin{equation}
\xi_{\pm}^{kl} (\theta)_{II,GI} = \frac{1}{2 \pi} \int_0^{\infty} d \ell~ \ell J_{0,4} (\ell \theta) C_{II,GI}^{k l} (\ell),
 \end{equation}
 \noindent
 where as before the ``+" (``-") component of $\xi_{\pm}^{kl}$ is evaluated with $J_{0(4)}$. 
 
In our cosmological parameter estimation, we marginalize over the interval $5.14<A<6.36$ with a flat prior, although we find that skipping this marginalization virtually creates no
difference in the final results.
One of the limitations of this luminosity-dependent nonlinear IA model is the narrow range of redshift used by Joachimi et al. (2011) to determine the parameter values. Because the DLS is a deep survey reaching $\langle z \rangle \sim1.2$, the application of this model requires a significant extrapolation. Since the $\left <L \right >/L_0$ value at $z\gtrsim1.2$ rises steeply (also depending on the details of the luminosity function shape), it may appear that the impact of the assumption on a fixed luminosity function slope is non-negligible. However, the overall effect of the depth $r_{\mathrm{lim}}\sim27$ of the DLS in fact mitigates the IA substantially because the fraction of LRGs at a fixed redshift decreases; one should remember that Figure C.1 of Joachimi et al. (2011) is based on the survey with $r_{\mathrm{lim}}=25$. In \textsection\ref{section_analysis}, we discuss the impact of the IA models on the parameter constraints.

Another potentially interesting issue, which is however neglected in Joachimi et al. (2011), is the IA contribution from satellite galaxies. 
The only statistically significant observational constraints are for an LRG selected sample in Singh et al. (2014). For galaxy populations more similar to those in DLS, the satellite intrinsic alignment measurements in the literature are all consistent with null detections (Hung \& Ebeling 2012, Schneider et al. 2013, Chisari et al. 2014).
Although the absence of satellite IAs is not likely to affect the current DLS tomographic study, the impact of the satellite galaxy IA on future cosmic shear surveys, where the interpretation is no longer limited by statistics, remains to be seen.   We summarize tests of the impact of IA models on our results in \textsection\ref{section_IA_impact}.
    
\section{ANALYSIS} \label{section_analysis}
\subsection{DLS Tomographic Bins}
The merit of tomographic cosmic shear analysis over a two-dimensional study  is the explicit utilization of the redshift-dependence of cosmic shear signals (due to both redshift-dependent structure growth and geometry). Of course, at the same time any redshift-dependent systematics become more important and affect the interpretation. One of the important questions in tomographic studies is the number of tomographic bins. Finding an answer to this question is non-trivial because there are many parameters to be considered such as the depth of the survey, mitigation of IA signals, photometric redshift uncertainty, stability of covariance matrix, type of information to be constrained, etc.  

Bridle \& King (2007) studied the issue in the presence of IA systematics and concluded that future surveys need as many as eight bins 
in order to utilize$\mytilde80$\% of the available information, assuming that we can control photometric redshift uncertainties reasonably well ($\delta z/(1+z)\sim0.05$).  On the other hand, Ma, Hu, \& Huterer (2006) estimated that saturation occurs quickly beyond $N_t\sim2$ and argued that $N_t\sim5$ will be a sufficient number for future surveys targeted for dark energy. The difference between the two studies may arise from the additional consideration of IA in Bridle \& King (2007), implying that when IA effects are considered, more tomographic bins may be preferred.
Interestingly, the two-tomographic bin analysis of the CFHTLenS data by Benjamin et al. (2013) shows no gain in the cosmological leverage over a single bin (two-dimensional) study of Kilbinger et al. (2013) whereas the six-bin analysis of Heymans et al. (2013) demonstrates that the improvement in parameter constraint is significant, shrinking the 1 $\sigma$ contour area in the $\Omega_m-\sigma_8$ plane by more than a factor of two.

In this study, we choose to set up five redshift bins. Although in this decision we take into account  the DLS depth, our photometric redshift uncertainty, and the feasibility of covariance matrix construction from existing $N$-body simulation data, we do not fully explore the issue regarding the optimal number of bins versus the constraining power.  One outstanding factor limiting this exploration is the evaluation of the covariance matrix, which requires expensive ensembles of ray-tracing
through $N$-body simulations. Nevertheless, 
because of the broad width of the lensing kernel in $z$, we 
doubt that any such optimization will make a significant difference in figure of merit.

\begin{figure}
\includegraphics[width=8cm]{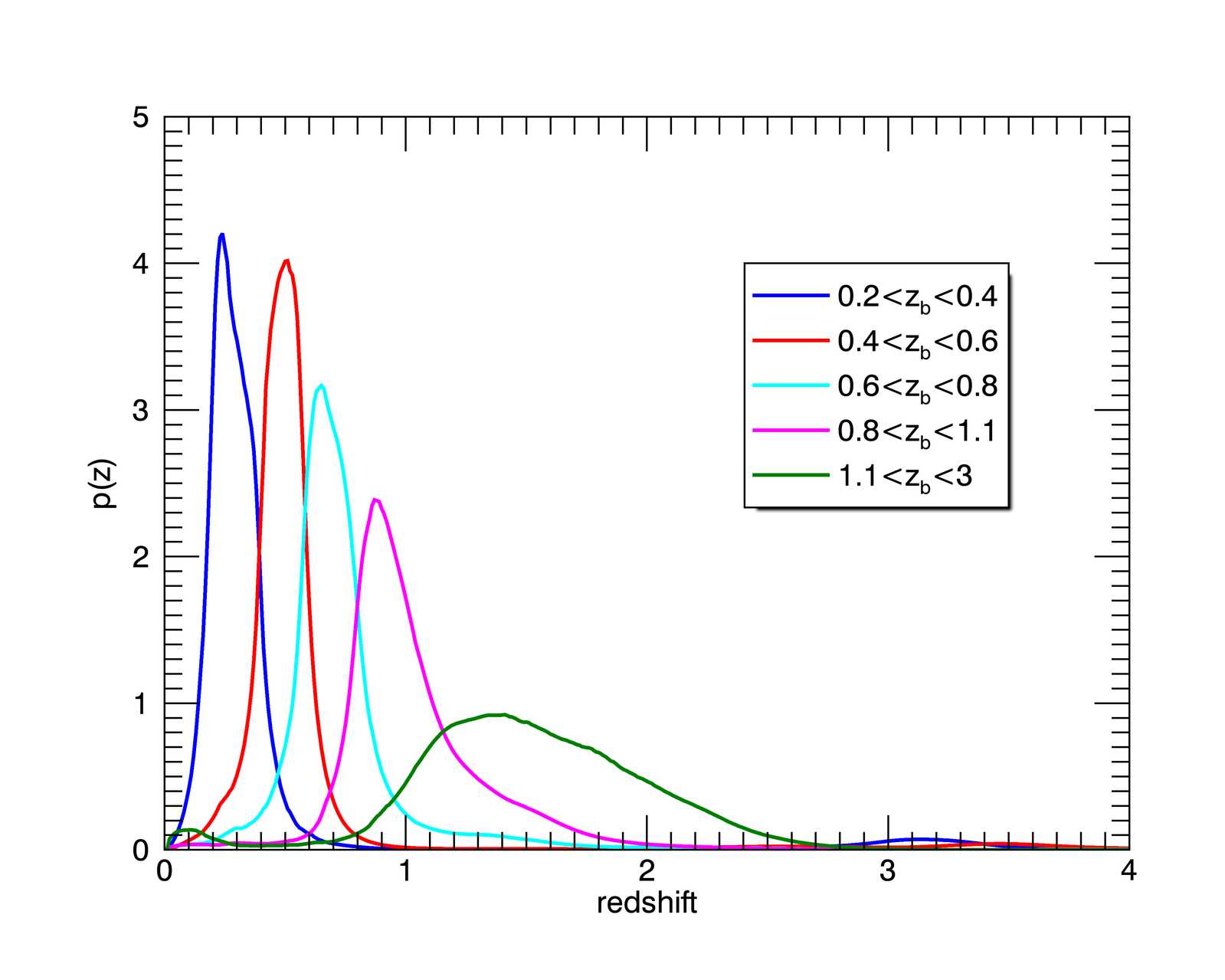}
\caption{Redshift distributions in the five DLS tomographic bins. Displayed are the stacked probability $p(z)$ (estimated by BPZ) of sources in each bin. The curves are normalized to unity integrated area. Table~1 lists the total number of sources in each redshift bin and more statistical details.}
\label{fig_p_of_z}
\end{figure}

Figure~\ref{fig_p_of_z} displays the redshift distributions of our source galaxies in the five DLS tomographic bins. We use  single-point estimates ($z_b$) from BPZ to determine which bin individual sources belong to; the photometric redshift intervals are shown in Table~1. The displayed curves are
the stacked probability $p(z)$ (estimated by BPZ) of sources in each bin.
We discard sources if their integrated probability within that interval is less than 20\% for bins 1-4 and less than 30\% for bin 5. 
By imposing this condition, we lose $\mytilde3$\% of source galaxies.
As discussed in J13, Abrahamse et al. (2011),  and Wittman (2009), the use of the full probability $p(z)$ instead of histograms constructed from single-point estimates have several advantages including
mitigation of the impact of catastrophic errors, reduction of bias, assignment of proper smoothing kernel due to photo-$z$ uncertainty, etc. Thus, we use the stacked $p(z)$ curves in Figure~\ref{fig_p_of_z} in our evaluation of the likelihood. Table~1 lists some details of the statistical properties in individual bins. 

\begin{deluxetable}{cccccc}
\tabletypesize{\scriptsize}
\tablecaption{Statistical properties of galaxies in DLS tomographic bins.}
\tablenum{1}
\tablehead{\colhead{Bin} &  \colhead{Range in $z_b$} & \colhead{$\left <z_b \right>$}   & \colhead{$\left < p(z) \right >$} &  \colhead{$n_{raw}$} &  \colhead{$n_{\mathrm{eff}}$}  }
\tablewidth{0pt}
\startdata
1   & 0.2--0.4 & 0.29 & 0.45 & 223,678 & 205,302  \\
2   & 0.4--0.6 & 0.48 & 0.62 & 273,441 & 232,802  \\
3   & 0.6--0.8 & 0.68 & 0.75 & 235,662 & 196,733  \\
4   & 0.8--1.1 & 0.93 & 1.04 & 234,413 & 189,400  \\
5   & 1.1--3.0   & 1.50 & 1.51 & 335,465 & 243,864 
\enddata
\tablecomments{A single point estimate $z_b$ is used to define each galaxy's membership within the five tomographic bins. However, we use the probability distribution $p(z)$ to predict their cosmic shear signals. Note that $\left < z_b \right >$ is systematically lower than $\left < p(z) \right >$, and the difference is larger for low redshifts. On the other hand, the ratio $n_{\mathrm{eff}}/n_{\mathrm{raw}}$ decreases with redshift simply because shape measurements at higher redshifts are noisier.}
\end{deluxetable}

\subsection{DLS Cosmic Shear Measurements} \label{section_measurement}

\begin{figure*}
\includegraphics[width=17cm,trim=2cm 0cm 2cm 0cm]{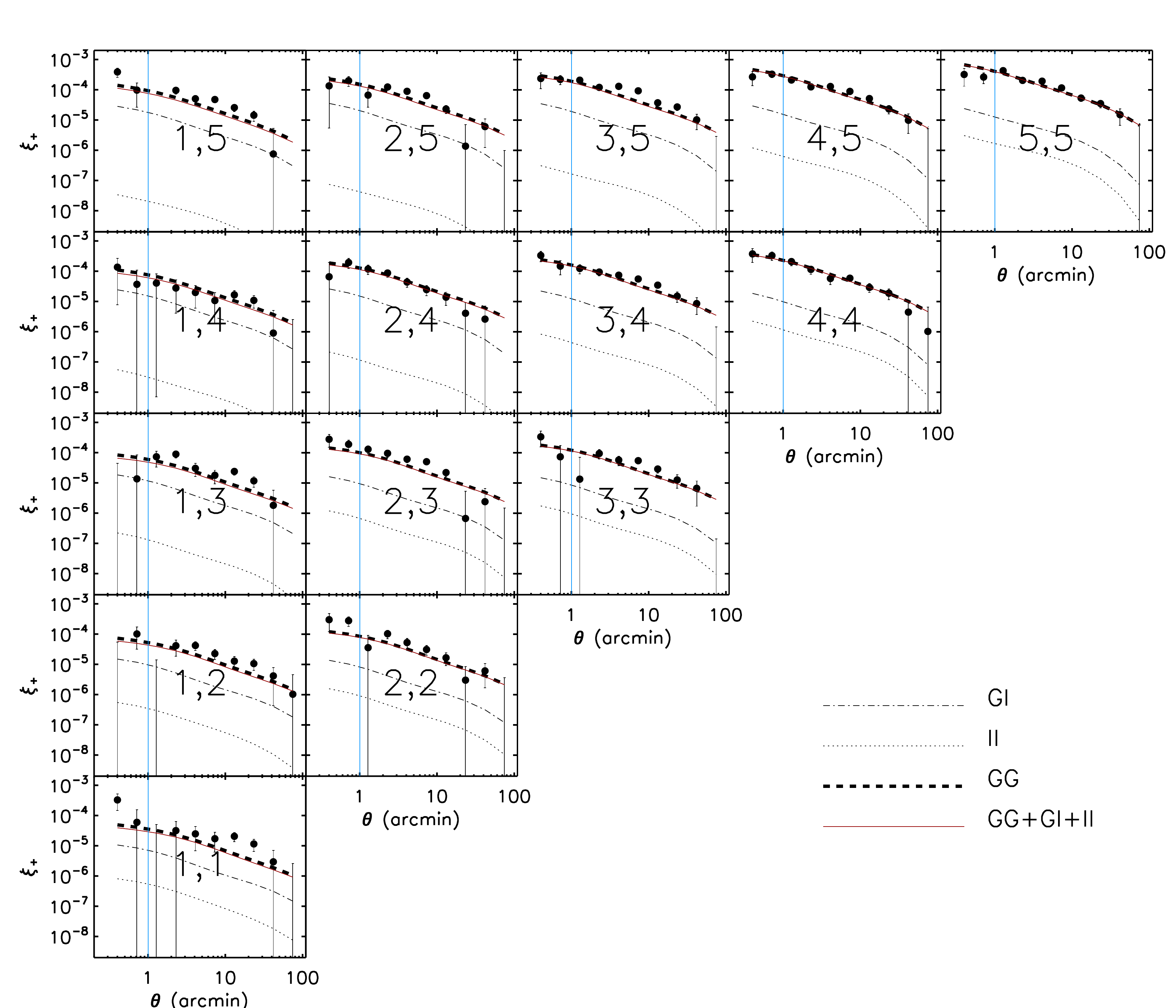}
\centering
\caption{Auto- and cross-correlation functions from the five DLS tomographic bins. 
We display cosmic shear (GG), shear-intrinsic alignment (GI), intrinsic-intrinsic alignment (II), and total (GG+GI+II) signals evaluated at the best-fit cosmology from the Planck2015-CMB result.
For constraining cosmological parameters, we use the data
points only at $\theta>1\arcmin$ to mitigate baryonic effects (solid blue line).
Note that because the GI signals are negative, we artificially flip their signs here to show them in these log-log plots. The two numbers $k,l$ displayed in each panel represent the tomographic bin combination. The contamination of intrinsic alignment to the DLS cosmic shear measurement is mostly negligible. Even when the lowest redshift ($z=0.29$) bin is involved, the amount of the shift in the correlation functions due to the correction is a small ($\lesssim10$\%) fraction of their statistical uncertainties. We discuss their impact on our cosmological parameter constraints in
\textsection\ref{section_IA_impact}. As demonstrated in \textsection\ref{section_comparison_with_planck2015}, the DLS cosmic shear signals are fully consistent with the Planck2015-CMB results.
}
\label{fig_xi_p_all}
\end{figure*}

Our 5-bin tomographic data consist of a total of 30 auto- and cross-correlation functions $\xi^{k,l}_{+,-}$.
In Figure~\ref{fig_xi_p_all}, we display the 15 $\xi^{k,l}_+$ functions along with the predicted signals from the cosmic shear and intrinsic alignments. Although not shown here, the signal from each tomographic bin passes the standard null test that we demonstrated in J13.
For the cosmic shear signal prediction, we assume the best-fit cosmological parameters published by Planck2015-CMB using the DLS photometric redshift distributions $p(z)$ (Figure~\ref{fig_p_of_z}). The IA signal is evaluated
using the Joachimi et al. (2011) model. The amplitude of the expected IA signal is small, and thus the contamination to the DLS cosmic shear is negligible according to this model.
Although Figure~\ref{fig_xi_p_all} shows the correlation functions evaluated at 10 logarithmically spaced
radial bins spanning $0.4\arcmin < \theta < 90\arcmin$,  to mitigate baryonic effects we
discard the two innermost data points at $\theta<1\arcmin$ when
constraining cosmological parameters (\textsection\ref{section_data_vector}). The resulting minimal separation $\theta\sim1.3\arcmin$ is 
similar to the value $\theta\sim1.5\arcmin$ in Heymans et al. (2013).

\begin{figure*}
\centering
\includegraphics[width=8.5cm,trim=1cm -0.5cm 0cm 0cm]{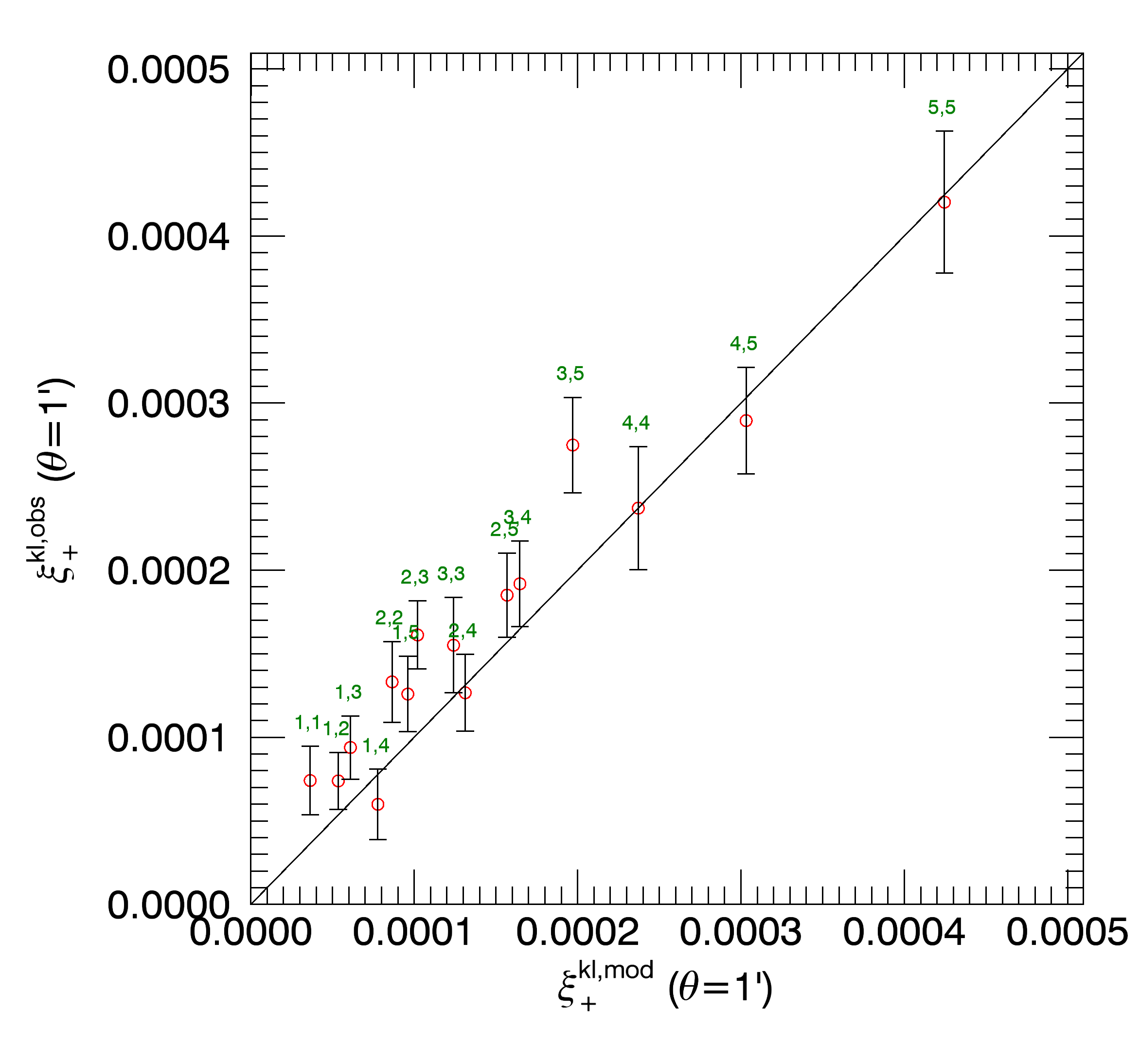}\includegraphics[width=8.5cm,trim=0cm -0.5cm 1cm 0cm]{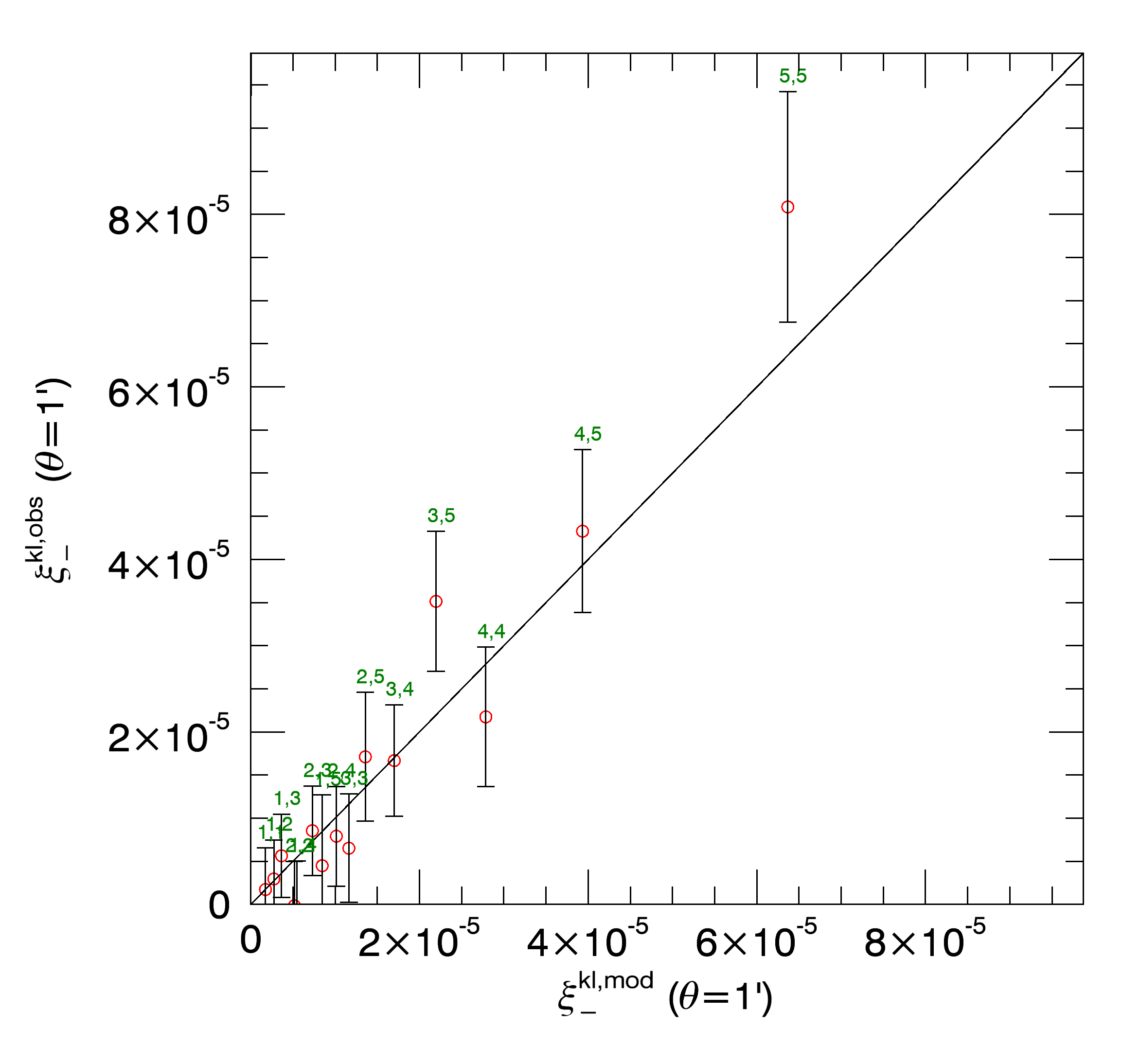}
\caption{Compressed representation of our DLS cosmic shear tomography. The 15 data points represent the signal amplitudes of all 15 correlation functions $\xi_{+,-}^{k,l}$ at $\theta=1\arcmin$. The $x$-axis $\xi_{+,-}^{k,l,mod}$ is the model prediction 
 at the reference cosmology (the Planck-CMB result) whereas the y-axis $\xi_{+,-}^{k,l,obs}$
 corresponds to the DLS observation (see text for the compression scheme). The solid line is just the equality, not a fit to the data. The error bars include the sample variance.}
\label{fig_compressed_xipm}
\end{figure*}

Figure~\ref{fig_xi_p_all} allows us to assess how the measured correlation functions compare to the predicted ones across all angular scales in detail. However, some authors prefer to compress the information further and visualize all the comparisons in a single plot (of course at the expense of some information loss). This is possible if we parametrize the difference between observation and prediction for each panel in Figure~\ref{fig_xi_p_all}. We follow the scheme proposed by Schrabback et al. (2010), who suggested to define the parameter as:
\begin{equation}
\alpha_{+,-}^{k,l} =\frac{ \xi_{+,-}^{k,l,obs} (\theta)  } {\xi_{+,-}^{k,l,mod} (\theta)},
\end{equation}
\noindent
where $\xi_{+,-}^{k,l,mod} (\theta)$ is the model prediction at the reference cosmology and 
$\xi_{+,-}^{k,l,obs} (\theta)=\alpha_{+,-}^{k,l} \xi_{+,-}^{k,l,mod}$ is the re-scaled prediction to best-fit the observation.
Since correlation functions are evaluated at multiple
angular scales, it is necessary to select a representative angular bin for visualization. 
In Figure~\ref{fig_compressed_xipm}, we display this compressed visualization for the choice $\theta=1\arcmin$.
This comparison shows that our DLS tomographic data follow a redshift-scaling relation and overall the result is consistent with the expectation for the reference cosmology. We note that the plot indicates that the observed $\xi_+$ signals
at the two lowest redshift bins are slightly higher than the prediction. However, because the measurements are correlated, the combined tension is only marginal at the $\mytilde1.5$ $\sigma$ level. If the Planck2015-CMB cosmological result is unbiased, this marginal tension may arise from any residual photometric redshift systematics at the lowest redshift bins, or our IA model may under-predict the level of II alignments. Neither issue can be resolved given the current data and theoretical understanding. We note that a similar level of signal excess at low-$z$ bins is also present in the CFHTLenS tomographic study (Heymans et al. 2013).

\section{COSMOLOGICAL PARAMETER CONSTRAINTS} \label{section_parameters}

\subsection{Data Vector, Signal Prediction, and Likelihood}
\label{section_data_vector}

Our data vector consists of 240 elements ($8\times2\times15$) ordered as follows:
\begin{equation}
\symvec{\xi} = \left [ \xi_{+}^{1,1},\xi_{-}^{1,1},\xi_{+}^{1,2},\xi_{-}^{1,2}, ..., ...,
\xi_{+}^{4,5}, \xi_{-}^{4,5}, \xi_{+}^{5,5}, \xi_{-}^{5,5} \right ].
\end{equation}
\noindent
In other words, the $\xi_{+,-}^{k,l}$ functions are arranged in a lexical order, and for
a given combination $(k,l)$, the ``+" components precede the ``-" components.
Within $\xi_{+,-}^{k,l}$, the correlation function values are listed in the ascending order of angle. One should be mindful of this ordering scheme when making use of our online DLS tomography cosmic shear data\footnote{http://dls.physics.ucdavis.edu}.

We follow the same procedure as our 2D paper in evaluating the predicted signals.
That is, we compute shear power spectrum using the modified transfer function of Eistenstein \& Hu (1998) that includes Baryonic Acoustic Oscillations (BAO) features and the Smith et al. (2003) HaloFit nonlinear power spectrum. As before, we do not
account for the baryonic contribution to the power on small scale and the reduced shear correction (Kilbinger 2010). The absence of the reduced shear correction 
may bias the $\sigma_8$ value by $\mytilde1$\% (Schrabback et al. 2010), which can be ignored in the current study, where the contribution from other errors (e.g., sample variance, shot noise, etc.) is much larger.
The baryonic effect may not be negligible, however. 
Many authors studied the effect through numerical simulations (e.g., Hearin \& Zentner 2009; Semboloni et al. 2011;
Semboloni et al. 2013).
Unfortunately, currently no convincing model exists. 
According to Eifler et al. (2014), different 
baryonic models can shift the 
cosmic shear power spectrum from $\mytilde2$\% to $\mytilde20$\% 
at $l\sim2000$ with respect to the dark matter only case. Also, the direction of the shift is model-dependent. Therefore, the only measure that we take in this study to mitigate the effect is to discard the signals at $\theta<1\arcmin$.

The likelihood function is given by:
\begin{equation}
\mathcal{L} =  \frac{1}   { (2 \pi)^{n/2} | \textbf{C} |^{1/2} }   \exp{  \left [ - 0.5  \left (\hat{\symvec{\xi}} - \symvec{\xi} \right )^{\mbox{\tt T}}  \textbf{C}^{-1}     \left ( \hat{\symvec{\xi}} - \symvec{\xi} \right )
        \right ] },
\end{equation}
\noindent
where \symvec{\xi} is the model prediction for a given set of cosmological parameters,
\textbf{C} is the covariance matrix, and $n$ is the number of elements in the data vector $\hat{\symvec{\xi}}$.

The covariance $\textbf{C}$ is decomposed as
$\textbf{C} = \textbf{C}_n + \textbf{C}_s + \textbf{C}_B + \textbf{C}_{\epsilon}$, where $\textbf{C}_n$ is the statistical noise, $\textbf{C}_B$
is the residual systematic error,  $\textbf{C}_s$ is the sample variance, and $\textbf{C}_{\epsilon}$ is a
cross-term between shape noise and shear correlations (Joachimi et al.  2008).
Because the sample variance is cosmology-dependent and also a significant
contributor to the total error budget, it is logical to let this component of the covariance vary in parameter estimation if possible. This cosmology-dependent 
covariance (CDC) scheme can be easily implemented if we are only concerned about cosmic shear signals on large angular scales ($>10\arcmin$), where the
covariances are virtually Gaussian. Gaussian covariances are readily obtained by simple analytic integration:
\begin{equation}
\mbox{Cov}[\xi_+ (\theta), \xi_+ (\theta^{\prime}) ] = \frac{1}{\pi \Omega} \int l~ \mbox{d} l~ J_0 (l\theta)J_0 (l \theta^{\prime}) P_{\kappa} (l)^2,  \label{eqn_gauss_cov}
\end{equation}
\noindent
where $\Omega$ is the sky area of the survey.
However, the Gaussian covariance underestimates the variance in the non-linear regime 
($<10\arcmin$) by up to several factors. 
For estimation of the non-Gaussian covariance, often expensive numerical simulations followed by ray-tracing are used.
In our 2D analysis, we implemented this CDC scheme by first estimating non-Gaussian covariances from the Sato et al. (2011) $N$-body simulation data at a fiducial cosmology and then assuming that the ratio of this non-Gaussian to Gaussian covariances is constant across different cosmologies.

However, in the current DLS tomography with 15 auto- and cross-correlations, it is difficult to apply the same strategy not only because the computation is expensive, but also because the covariance estimator itself becomes noisier. 
Therefore, we decided not to implement the CDC scheme in the current tomographic study, although we acknowledge that it is worth investigating the impact of CDC in future studies. In our 2D analysis, we find that CDC improves parameter constraints in the
$\Omega_m$-$\sigma_8$ plane compared to the case without CDC. Although, the greatest improvement is along the degeneracy direction in the $\Omega_m$-$\sigma_8$ plane.

\subsection{Covariance Matrix from N-Body Simulations}
We compute estimates of the data covariance and its cosmic variance part from a large sample of simulated DLS-like fields created by ray-tracing through $N$-body simulations of structure formation. We create a total of 2048 mock DLS fields of
2$\times$2 deg$^2$, 256 fields from each of 8 `fiducial' 
{\tt zHORIZON} simulations (Smith 2009).
These simulations employ {\tt GADGET-2} (Springel 2005) with
$N=750^3$ particles to follow the evolution of cosmic structure in a cubic region of 1500 $h^{-1}$ Mpc comoving side length in a flat $\Lambda$CDM
universe with $\Omega_m=0.25$, $\Omega_\Lambda = 0.75$, $\Omega_b=0.04$, $\sigma_8=0.8$,
and $n_s=1$. From the $N$-body
outputs, we create lensing lightcones
by backward ray-tracing (see Hilbert et al. 2007, 2009, 2012, for details of the method). The lightcones are then populated with source galaxies with random sky positions in a way that reproduces the number densities and redshift distributions of the actual DLS tomographic bins.

In each resulting simulated DLS field, the shear correlation functions $\xi_{+,-}^{k,l}$ are measured in 8 radial bins following the DLS radial bin setup. The measurements are performed both without and with adding shape noise.
The cosmic variance part of the data covariance for a single DLS field is then estimated from the sample variance of the shape noise-free measurements in the simulated fields (Hilbert et al. 2011). The full data covariance is estimated from the simulations including shape noise. The covariance for the full five-field survey is then obtained by rescaling the covariance estimate for a single field. Assuming that the 2048 simulated DLS fields are statistically independent, we obtain a very small relative statistical uncertainty $\delta\textbf{C/C}\approx3$ \% for these covariance estimates (Taylor \& Joachimi 2014).

We also create mock DLS fields by populating the 2048 simulated lightcones with source galaxies at the same angular positions as in the DLS fields. The resulting mock fields thus not only feature the same source redshift distributions but also the same angular source distributions as the actual DLS fields. The simulated individual DLS fields are then combined into 409 mock realizations of the full five-field DLS. From the shear correlation functions measured in these mock realizations, we again obtain estimates for the cosmic variance part of the DLS shear tomography data covariance, this time including effects of source clustering and gaps due to masked areas. Since the number of realizations for the full DLS is smaller by a factor five than the number of realizations for a single field, we estimate a larger relative statistical uncertainty $\delta \textbf{C/C}\approx10$ \% for these covariance estimates. Since we do not find any significant differences to the covariance estimates from the first method assuming uniform spatial distribution of sources, we only use those in the subsequent analysis.

\begin{figure}
\includegraphics[width=8.5cm,trim=1.2cm -1cm 0.8cm 0cm]{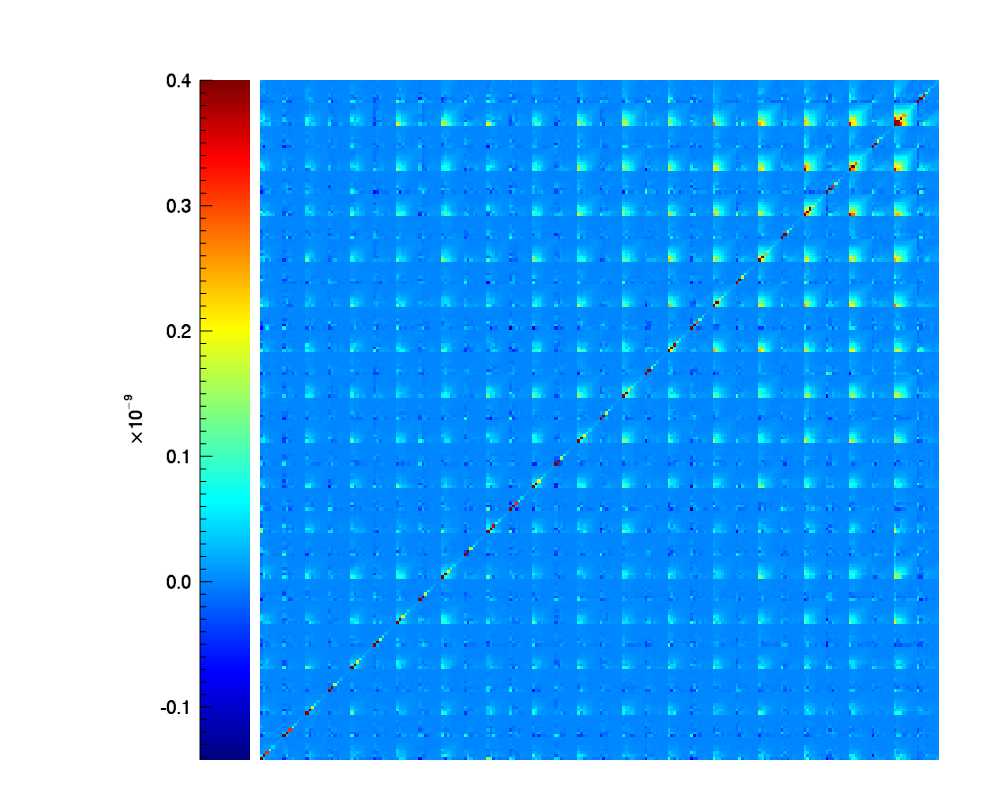}
\caption{Covariance matrix of the DLS tomography. We show the total covariance, which combines the sample variance, shape noise, mixed component, and masking effects. The dimension is 300$\times$300. The element ordering scheme is consistent with the data vector defined in \textsection\ref{section_data_vector}. Although the covariance spans from -$1.8\times10^{-9}$ to $+3.6\times10^{-8}$, we truncate the dynamic range as shown in order to highlight low-contrast structures.}
\end{figure}

\subsection{DLS-only Constraints} \label{section_DLS_only}

We consider only the $\Lambda$CDM cosmology for the DLS-only constraint. We let the Hubble constant $h$ have a flat prior between $0.6<h<0.8$. This range is a fair choice because the interval encompasses the 3-$\sigma$ range of both WMAP9 ($H_0=70.0\pm2.2$; Hinshaw et al. 2013) and Planck2015-CMB results ($H_0=67.48\pm0.98$) and also the Riess et al. (2011) Supernovae+Cepheid result ($H_0=73.8\pm2.4$).
Similarly, for the baryon fraction $\Omega_b$ and the spectral index $n_s$, we choose the intervals $0.03<\Omega_b<0.06$ and $0.92<n_s<1.02$ with flat priors, which span the $3\sigma$ ranges established by both WMAP9 and Planck2015-CMB.
We remind readers that cosmic shear is not sensitive to the above three parameters and enlarging the prior intervals do not significantly degrade the constraining power. The remaining two parameters $\Omega_m$ and $\sigma_8$ most sensitively affect cosmic shear signals, and we select large intervals ($0.01 < \Omega_m < 1.0$ and $0.1< \sigma_8 <1.2$).

\begin{figure}
\includegraphics[width=8cm,trim=0.5cm 0cm 1.5cm 0cm]{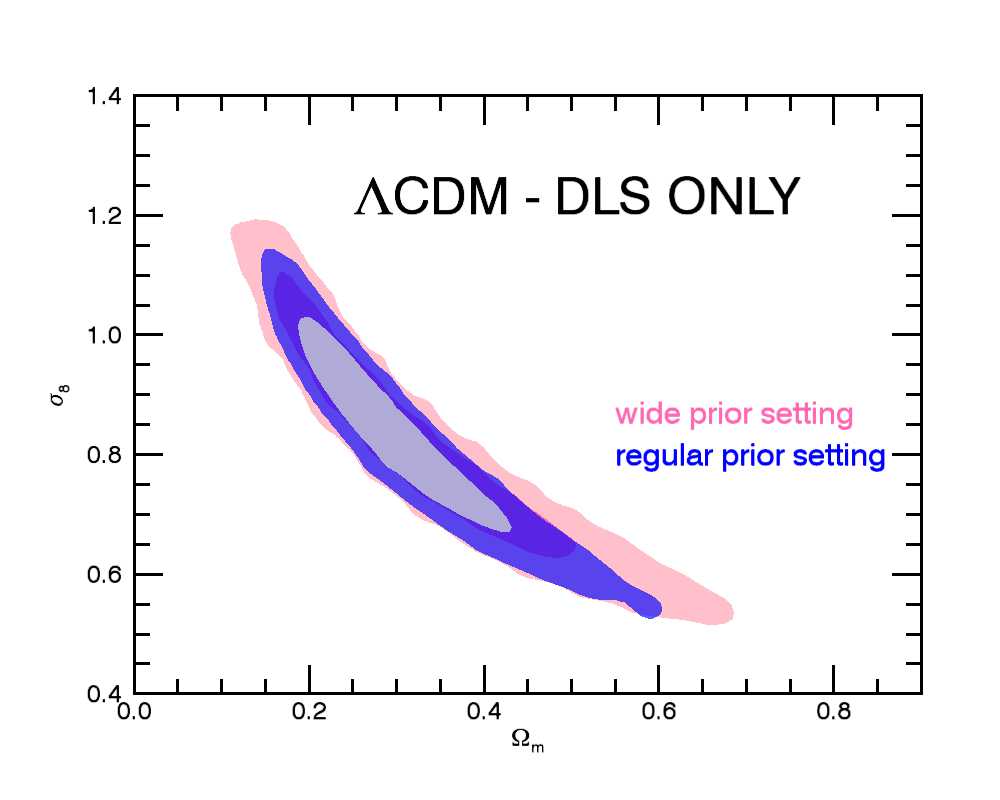}
\caption{``DLS-ONLY" constraints on $\Omega_m$ and $\sigma_8$ for $\Lambda$CDM.
The inner and outer contours represent 68\% and 95\% confidence regions, resp. Flat priors are used. For the ``regular" prior setting, we marginalize over the 
$0.6 < h < 0.8$, $0.92 < n_s < 1.02$, and $0.03 < \Omega_b < 0.06$ intervals, which bracket the 3$\sigma$ ranges constrained by previous CMB or SNIa+Cepheid studies. The ``wide" prior setting refers to the intervals: $0.4 < h < 1.2$, $0.7 < n_s < 1.2$, and $0 <\Omega_b< 0.1$, which are adopted in the CFHTLenS studies.
}
\label{fig_omega_m_vs_sigma_8_DLS_ONLY}
\end{figure}

Figure~\ref{fig_omega_m_vs_sigma_8_DLS_ONLY} displays the resulting ``banana" contours in the $\Omega_m$-$\sigma_8$ plane. Also shown is the result after we increase the prior intervals of $\Omega_b$, $n_s$, and $h$ to the same levels as is done in the CFHTLenS analysis (Kilbinger et al. 2013). Although our DLS tomographic cosmic shear does not lift the $\Omega_m$-$\sigma_8$ degeneracy substantially, the constraint in the direction perpendicular to the degeneracy (i.e., the width of the ``banana") is tight. Often, in cosmic shear studies this constraint in the $\Omega_m - \sigma_8$ plane is expressed in terms of $\sigma_8 \Omega_m^a$. Choosing $\Omega_m\equiv0.3$ as our pivot value, we obtain 
\begin{equation}
\sigma_8 \left ( \frac{\Omega_m}{0.3} \right)^{0.50} = 0.818_{-0.026}^{+0.034}. 
\end{equation}

\begin{figure*}
\includegraphics[width=15cm,trim=0.5cm 0cm 1.5cm 0cm]{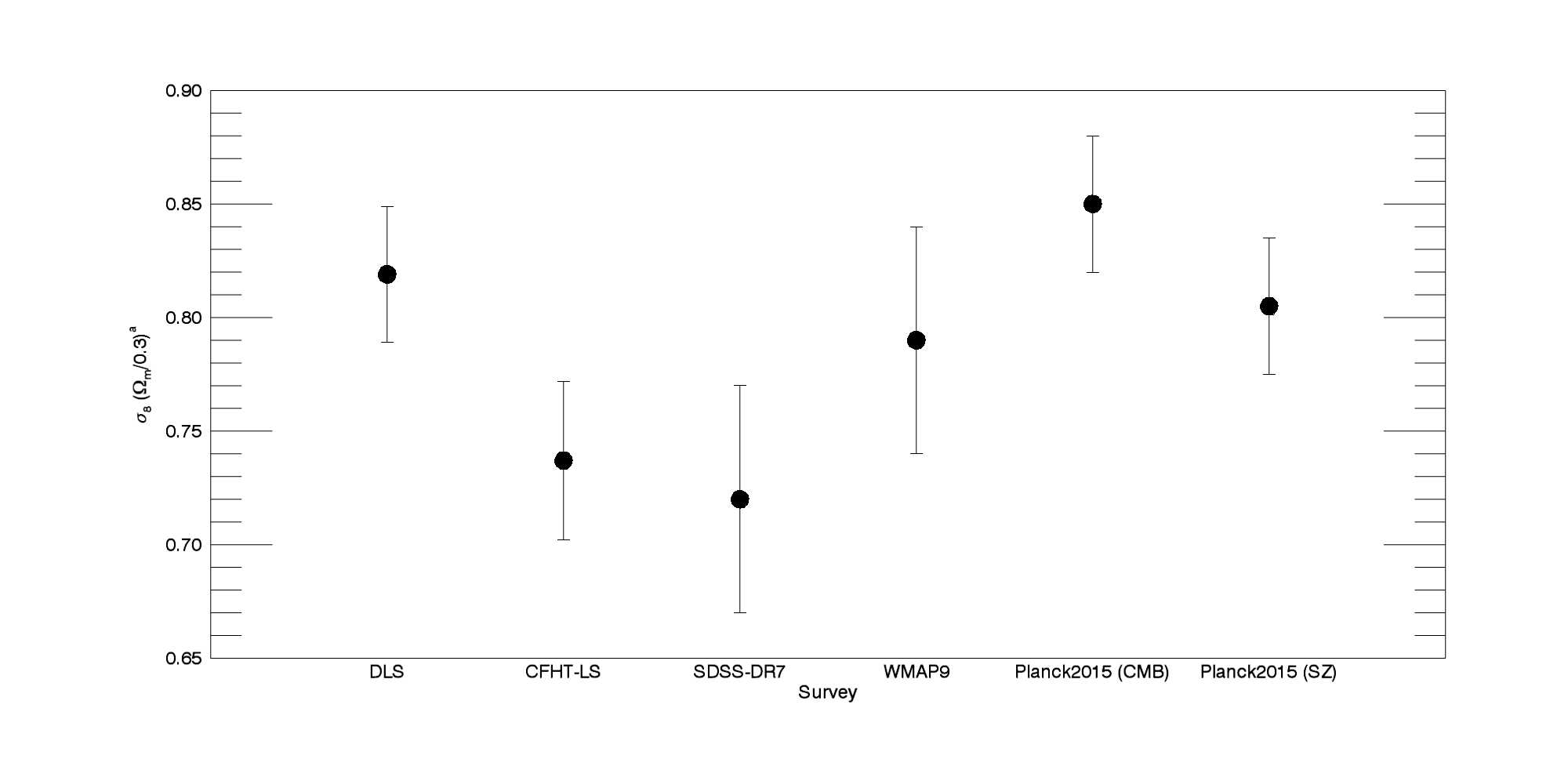}
\caption{Comparison of the constraint $\sigma_8 (\Omega_m/0.3)^a$ from different surveys normalized to $\Omega_m\equiv0.3$. Only a flat $\Lambda$CDM cosmology is considered.
We select studies
whose errors on $\sigma_8 (\Omega_m/0.3)^a$ are $\lesssim0.05$. If more than one study is available for a given survey, we quote the value from the most recent one (see text).
}
\label{fig_sigma_8_comparison}
\end{figure*}

The $\sigma_8 \Omega_m^a$ parametrization enables an
easy comparison of results from different studies.
According to the compilation of Kilbinger (2014), there are about
25 papers reporting on the $\sigma_8 \Omega_m^a$ constraint to date. 
We select studies whose errors on $\sigma_8 (\Omega_m/0.3)^a$ are $\lesssim0.05$. If more than one study is available for a given survey, we quote the value from the most recent one. In Figure~\ref{fig_sigma_8_comparison}, we compare the DLS result with
those from SDSS-DR 7 (Mandelbaum et al. 2013), CFHTLenS (Heymans et al. 2013), WMAP9 (Hinshaw et al. 2013), Planck2015-CMB (Planck Collaboration et al. 2015a), and Planck2015-SZ (Planck Collaboration et al. 2015b).
The result from Planck2015-SZ depends on the mass bias prior $1-b$ and the external data. We choose the result from the joint constraint with BAO using the value $1-b=0.688\pm0.072$ of von der Linden et al. (2014).

An interesting $\gtrsim2~\sigma$ tension is present between the Planck2015-CMB and the CFHTLenS results as also noted in previous studies (e.g., Planck Collaboration et al. 2015a; MacCrann et al. 2015). It is worth noting that the CFHTLenS data have been analyzed in several studies with slightly different techniques (e.g., Fu et al. 2014; Benjamin et al. 2013; Kilbinger et al. 2013; Heymans et al. 2013), which all provide highly consistent results. This low-normalization cosmology is also favored by Mandelbaum et al. (2013), who combine galaxy-galaxy lensing and galaxy clustering signals from SDSS-DR 7. 
If this tension between weak-lensing and CMB studies persists, the discrepancy may be interpreted as indicating some incompleteness in our understanding (e.g., MacCrann et al. 2015); some may also regard the difference between
Planck2015-CMB and -SZ results as supporting this low-$z$ vs. high-$z$ tension.
In this light, the DLS result is intriguing because the survey provides
one of the tightest constraints and is independent in its design and analysis method.
As shown Figure~\ref{fig_sigma_8_comparison}, the  $\sigma_8 \Omega_m^a$ constraint from DLS is consistent with those from both WMAP9 and Planck2015. Therefore, as far as the DLS result is concerned, 
it is premature to argue that new physics is required to resolve this low-$z$ vs. high-$z$ tension.

\subsection{Joint Probes with External Data}
We consider both $\Lambda$CDM and $w$CDM models with and without the curvature constraint. The flat $\Lambda$CDM model is
our baseline model and is described by the following five parameters: $\Omega_m$, $\sigma_8$, $\Omega_b$, $n_s$, and $h$. The flat $w$CDM extension requires one additional parameter $w$, which characterizes the equation of state parameter $w=p/\rho$. The parameter $\Omega_{\Lambda}$ is added when we relax the curvature constraint $\Omega_k \equiv 1-\Omega_m-\Omega_{\Lambda} \equiv 0$.

For external data, we use BAO, CMB and SNIa data and provide their details as follows.
We combine the Baryonic Acoustic Oscillation (BAO) results 
published by Anderson et al. (2012), Padmanabhan et al. (2012), Beutler et al. (2011), and Blake et al. (2012). These results were derived from the 6dFGS (Johns et al. 2004), SDSS-DR7, SDSS-DR9, and WiggleZ surveys and were also used by WMAP9 in their joint cosmological parameter constraint. Table 2 summarizes their effective redshifts and measurements on $D_V(z)/r_s$, where $r_s$ is
the sound horizon distance, and $D_V(z)$ is the distance measure at $z$ defined as:
\begin{equation}
D_V(z) = \left [ (1+z)^2 D_A^2 (z) \frac{cz}{H(z)} \right ]^{1/3}. \label{eqn_dv}
\end{equation}
\noindent
In equation~\ref{eqn_dv}, $D_A(z)$ is an angular diameter distance to the redshift $z$.
We use the covariances between the last three measurements in Table 2 published in Blake et al. (2012).

\begin{deluxetable}{lccl}
\tabletypesize{\scriptsize}
\tablecaption{BAO measurements used in the current joint constraint.}
\tablenum{2}
\tablehead{\colhead{z} & \colhead{$D_V(z)/r_s$} &  \colhead{Survey} & \colhead{Reference}   }
\tablewidth{0pt}
\startdata
0.1 &  $2.98\pm0.27$ & 6dFGS & Beutler et al. (2011) \\
0.35 & $8.88\pm0.17$ & SDSS-DR7 & Padmanabhan et al. (2012) \\
0.57 & $13.67\pm0.22$ & SDSS-DR9 & Anderson et al. (2012) \\
0.44 & $10.92\pm3.67$ & WiggleZ & Blake et al. (2012) \\
0.60 & $13.77\pm5.94$ & WiggleZ & Blake et al. (2012) \\
0.73 & $16.89\pm9.15$ & WiggleZ & Blake et al. (2012) 
\enddata
\end{deluxetable}

For the cosmic microwave background, we use the Wilkinson Microwave Anisotropy Map 9-year result (Hinshaw et al. 2013; hereafter WMAP9)\footnote{Although we do not directly use the Planck2015-CMB result, we will present the comparisons of our joint probe results with those from Planck2015-CMB in \textsection\ref{section_comparison_with_planck2015}}.
WMAP9 update their previous results based on the final 9-year data with some revised calibrations, improving the average parameter uncertainty by $\mytilde10$\% compared to their 7-year results (Komatsu et al. 2011).

For supernova data, we utilize the Union2.1 catalog\footnote{available at
http://supernova.lbl.gov/Union} provided by Suzuki et al. (2012). The compilation contains 580 supernovae distance moduli within the $0.015<z<1.41$ range. The supernova $\chi^2$ function is given by
\begin{equation}
\chi_{SNIa}^2 = \sum_i \frac{ \left [ \mu_B(\alpha,\beta,M_B) - \mu(z,\Omega_m,\Omega_{\Lambda},w) \right ]^2 } {\sigma_{total}^2}, \label{eqn_sn_chi}
\end{equation}
\noindent
where the summation is performed over 580 supernovae. The distance modulus $\mu_B$ is a function of the rest-frame B-band magnitude $m_B$, the universal absolute SNIa magnitude, $M_B$, the shape of stretch parameter $s$, and the color $c$:
\begin{equation}
\mu_B=m_B - M_B + \alpha(s-1) - \beta c
\end{equation}
\noindent
where the linear response parameters $\alpha=0.1219$ and $\beta=2.4657$ are determined globally by fitting all 580 supernovae in Suzuki et al. (2012). The best-fit parameter $M_B$ is $M_B=-19.3082$ at $H_0=70~\mbox{km~s}^{-1}~\mbox{Mpc}^{-1}$ when the known systematics are included\footnote{For each set of the cosmological parameters, one must adjust the value by $5\log(h/0.7)$. Omitting this is equivalent to imposing a $H_0$ prior centered at $h=0.7$.}.
We refer readers to Suzuki et al. (2012) for the evaluation of the term $\sigma_{total}$ in the denominator of Equation~\ref{eqn_sn_chi}. In our analysis, we use the version that includes the systematics, the propagated errors in light curve fitting, and the external errors such as those arising from Galactic extinction correction and gravitational lensing. 

For the DLS+BAO joint probe, we use the {\tt CosmoPMC} package (Kilbinger 2009), which explores parameter space efficiently through importance sampling. However, whenever the WMAP9 results are needed, we directly employ
the WMAP9 chains\footnote{http://lambda.gsfc.nasa.gov} provided by the team and importance-sample our DLS likelihood function with them. A joint likelihood is evaluated by simply multiplying the likelihoods of different probes.  
In principle, the likelihoods for individual probes are not statistically independent because of shared cosmic structures. However, the potential volumes of overlap of our current surveys are small enough that we expect a joint likelihood constructed from a product of individual probe likelihoods to be an excellent approximation.

\subsubsection{Matter Density $\Omega_m$ and Normalization $\sigma_8$}
\begin{figure*}
\centering
\includegraphics[width=8.8cm,trim=0.5cm 0cm 0cm 0cm]{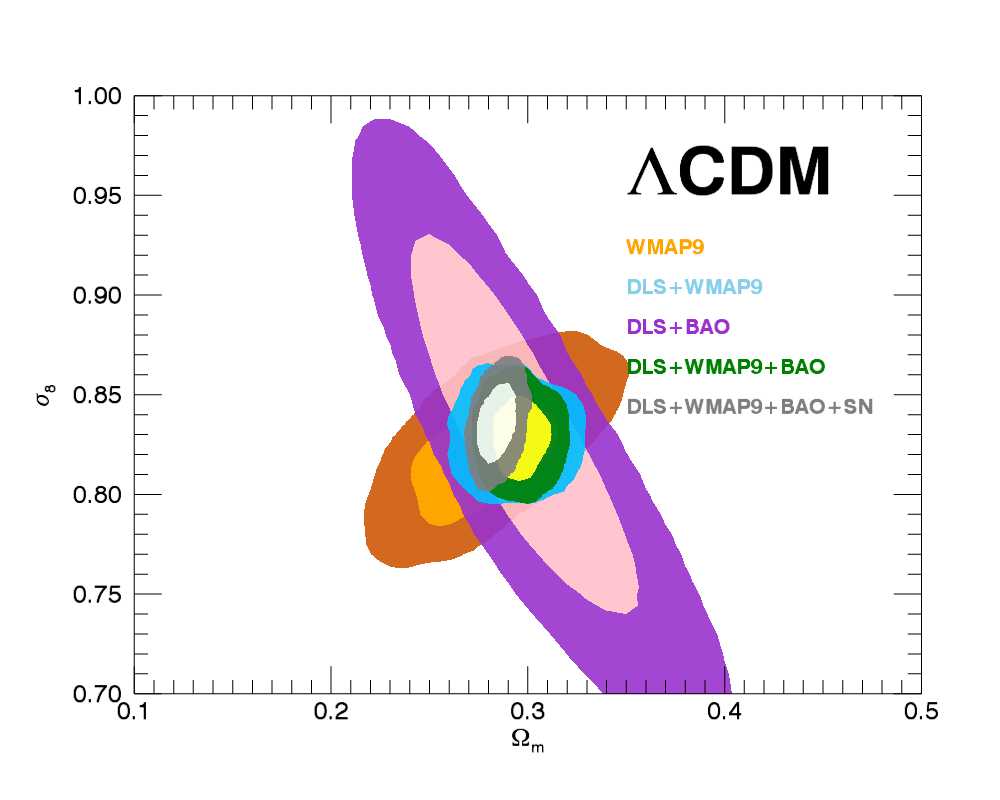}
\includegraphics[width=8.8cm,trim=0.5cm 0cm 0cm 0cm]{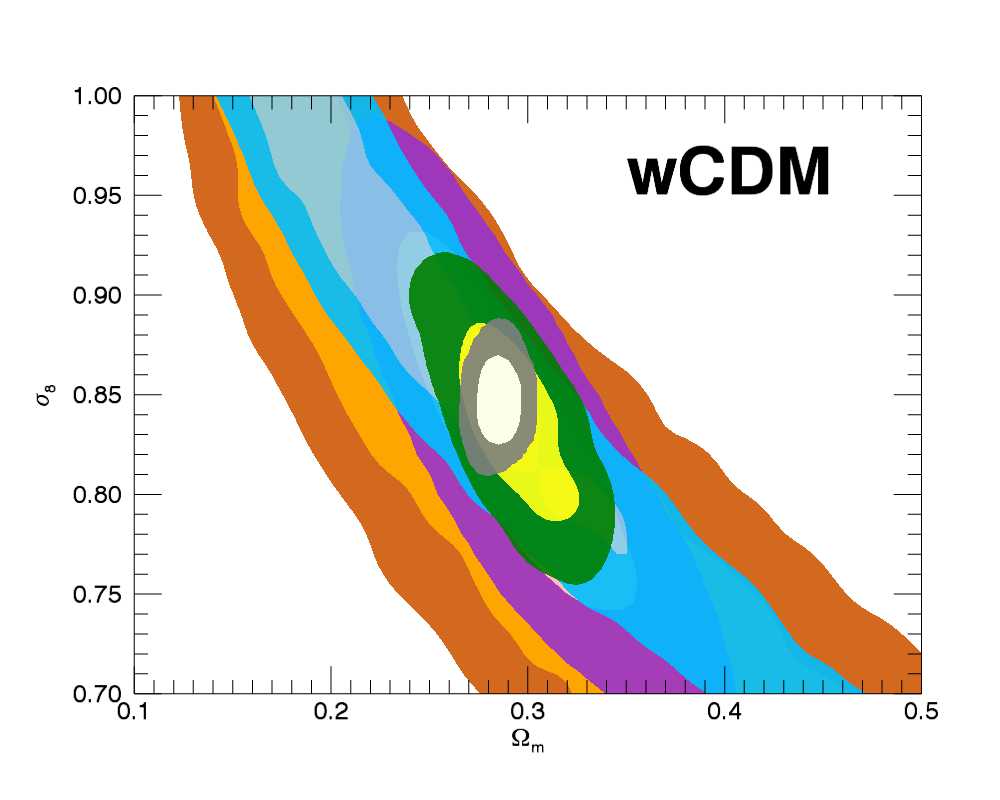}
\includegraphics[width=8.8cm,trim=0.5cm 0cm 0cm 0cm]{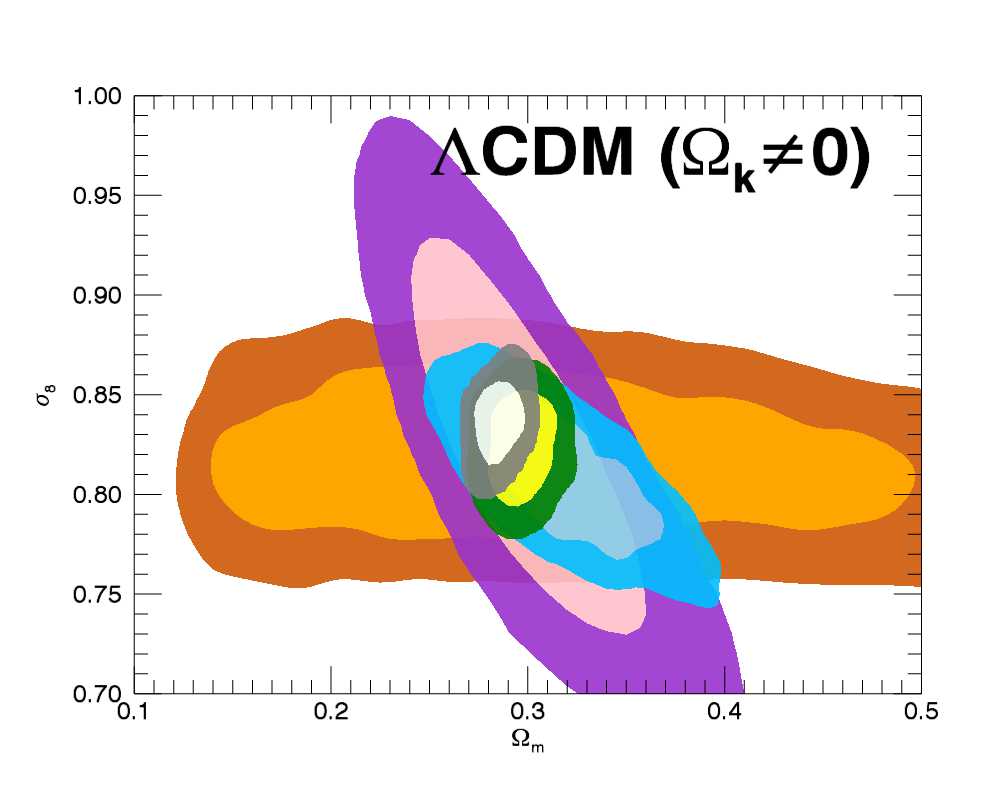}
\includegraphics[width=8.8cm,trim=0.5cm 0cm 0cm 0cm]{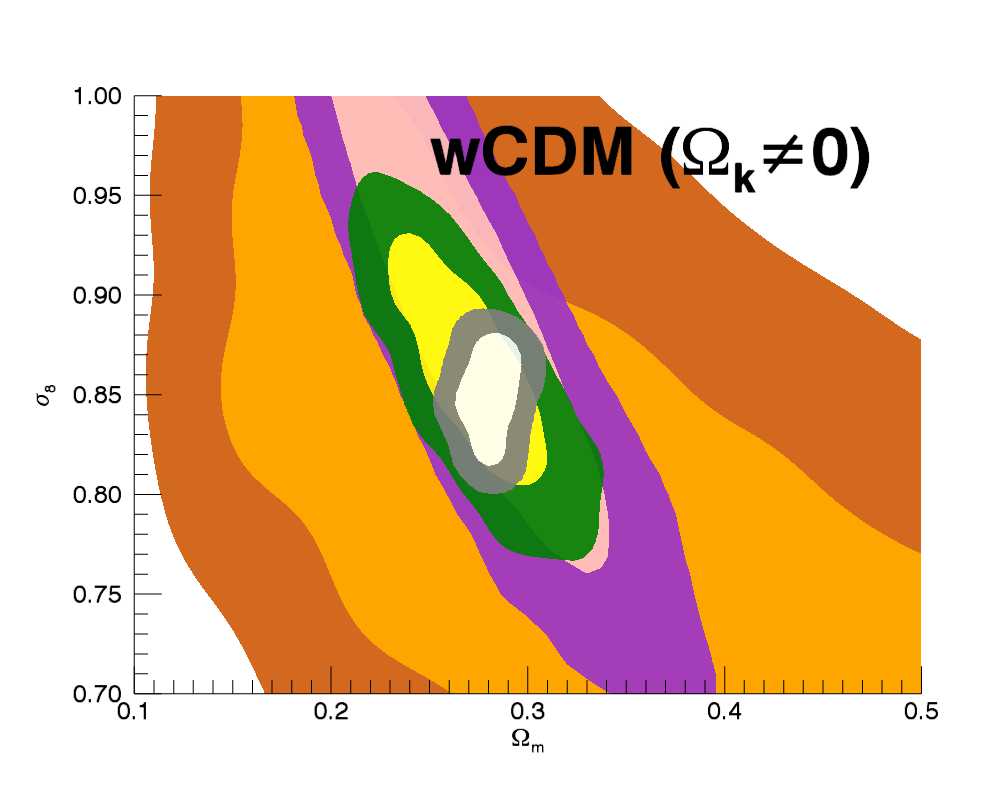}
\caption{Constraint on $\Omega_m$ and $\sigma_8$. The contribution of the DLS tomography is substantial in reducing the parameter uncertainties.}
\label{fig_omega_m_sigma_8_joint}
\end{figure*}

The degeneracy between $\Omega_m$ and $\sigma_8$ is lifted when
cosmic shear tomography is combined with other probes (Figure~\ref{fig_omega_m_sigma_8_joint}). Particularly, in the flat $\Lambda$CDM model, the degeneracy of the two parameters in CMB constraints is nearly orthogonal to that in cosmic shear. 
Compared to the case where WMAP9 alone is used, the addition of the current DLS tomography yields $\Omega_m=0.293_{-0.012}^{+0.014}$ and $\sigma_8=0.833_{-0.011}^{+0.018}$, shrinking
their 1-$\sigma$ uncertainties by $\mytilde48$\% and $\mytilde39$\%, respectively.
Adding the BAO distance prior further reduces their uncertainties by 
$\mytilde23$\% and $\mytilde28$\%, respectively.

If we relax the flatness constraint $\Omega_k\equiv 0$, the WMAP9 CMB alone  no longer constrain the value of $\Omega_m$ tightly ($0.19 < \Omega_m < 0.95$ for 95\% confidence). The roles of both the DLS cosmic shear
and BAO become critical in this case. When we use the DLS+WMAP9 joint probe,
we obtain $\Omega_m=0.315_{-0.024}^{+0.038}$ and $\sigma_8=0.805_{-0.025}^{+0.025}$. In
terms of the areas enclosed by 1-$\sigma$ contours in the $\Omega_m$-$\sigma_8$ plane,
the reduction is more than $\mytilde85$\%. When BAO is added, the results become
$\Omega_m=0.297_{-0.012}^{+0.011}$ and $\sigma_8=0.837_{-0.013}^{+0.022}$, in excellent agreement with the values evaluated for $\Lambda$CDM.

For a flat $w$CDM, the degeneracy between $\Omega_m$ and $\sigma_8$ constrained by WMAP9 alone is similar to the cosmic shear one and cannot distinguish the combination of a high normalization with a low matter density from the combination of a low normalization with a high matter density. Therefore, the addition of the DLS data is not effective in breaking the degeneracy, although the constraint orthogonal to this degeneracy is still significant and reduces the width of the ``banana" by $\mytilde60$\%. A meaningful constraint is obtained only when we combine the three probes, which gives $\Omega_m=0.290_{-0.017}^{+0.020}$ and $\sigma_8=0.845_{-0.039}^{+0.025}$,  in accordance with the case in $\Lambda$CDM.

Not surprisingly, the constraining power is substantially weakened for the non-flat $w$CDM. Nevertheless, we obtain $\Omega_m=0.269_{-0.024}^{+0.029}$ and $\sigma_8=0.853_{-0.033}^{+0.051}$ using all three probes jointly. We note that again these results are fully consistent with the values when different cosmologies are assumed.

Adding the supernovae data improves the constraints, and the improvement is particularly notable for the case of $w$CDM, where the $\Omega_M$-$\sigma_8$ degeneracy remains even with the DLS+WMAP9+BAO combination. For $\Lambda$CDM, the addition of the supernova data slightly ($\mytilde5$\%) decreases the matter density
(e.g., $\Omega_m=0.283_{-0.005}^{+0.007}$ for flat universe). 

\subsubsection{Hubble Constant $H_0$} \label{section_hubble}

\begin{figure*}
\centering
\includegraphics[width=8.8cm,trim=0.8cm 0cm 0cm 0cm]{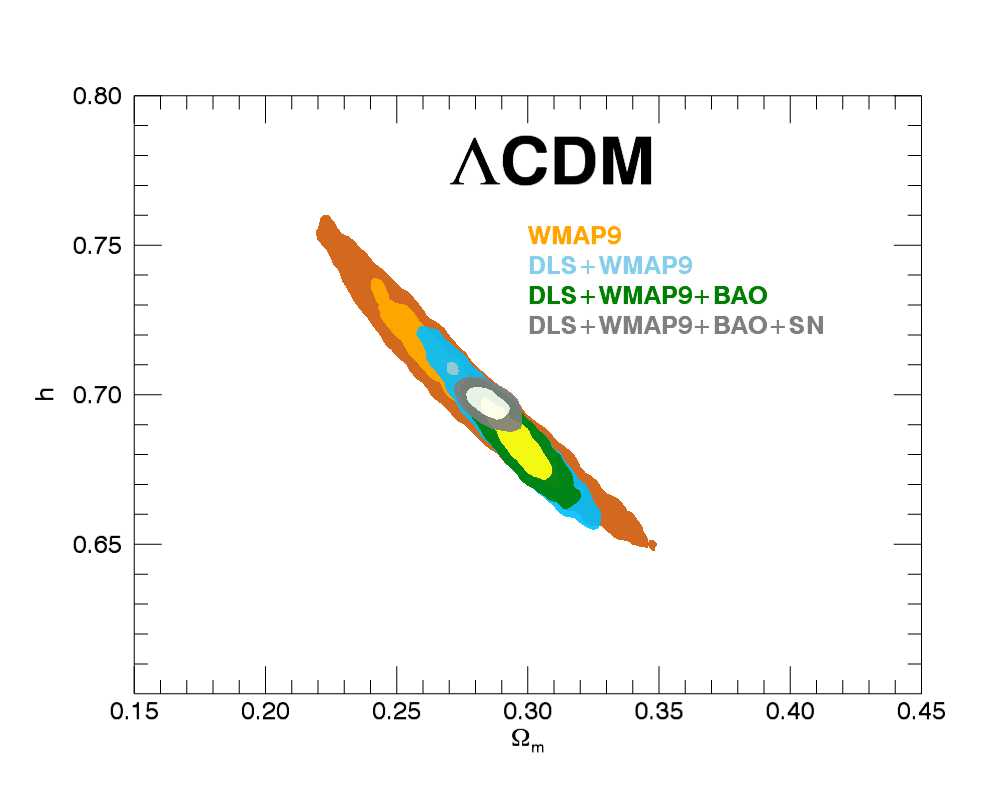}
\includegraphics[width=8.8cm,trim=0.8cm 0cm 0cm 0cm]{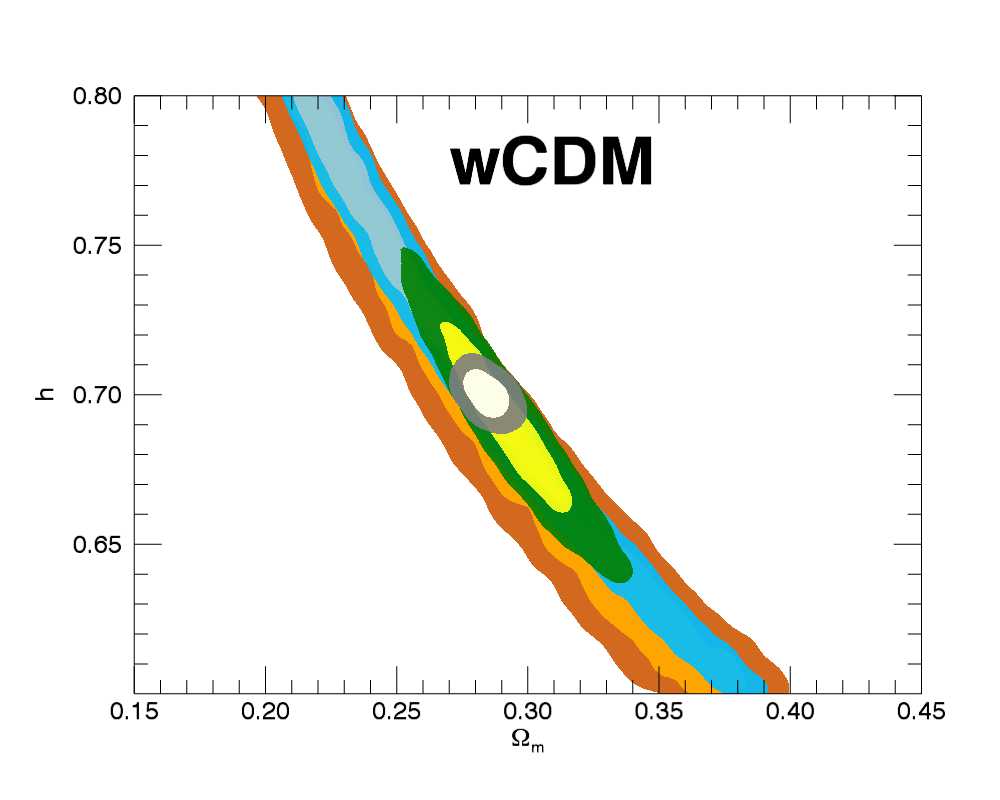}
\includegraphics[width=8.8cm,trim=0.8cm 0cm 0cm 0cm]{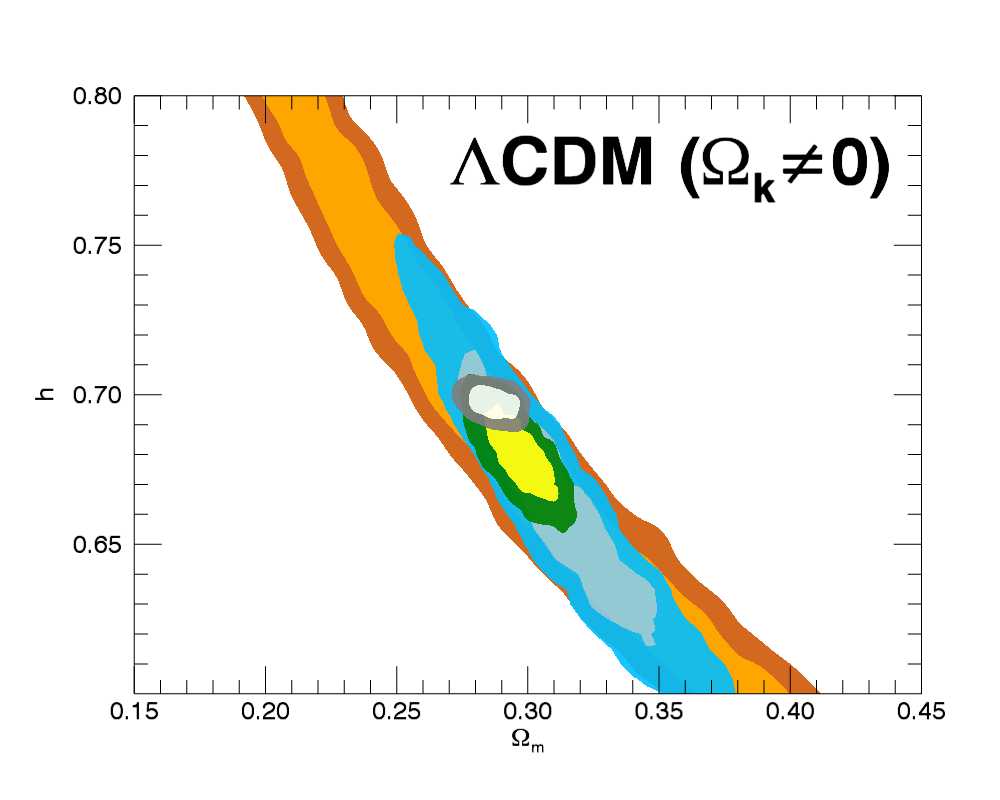}
\includegraphics[width=8.8cm,trim=0.8cm 0cm 0cm 0cm]{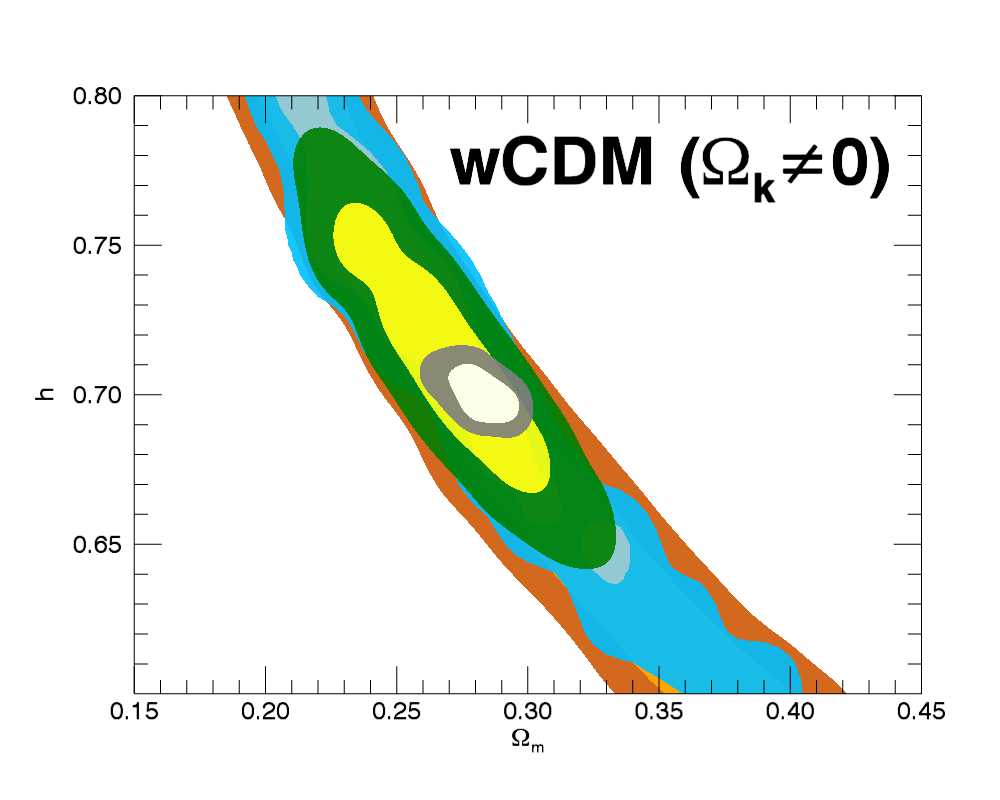}
\caption{Constraint on Hubble Constant $H_0$.}
\label{fig_hubble}
\end{figure*}

The DLS cosmic shear by itself does not provide useful constraints on the Hubble constant $H_0$ because the change in $H_0$ can trade with the change in the normalization of $\sigma_8 \Omega_m^a$.
In Figure~\ref{fig_hubble}, we illustrate this degeneracy with respect to the parameter $\Omega_m$.
However, as shown, when the DLS cosmic shear is used in conjunction with the other probes, the effect is significant. For $\Lambda$CDM, the DLS+WMAP9 joint analysis provides a tight constraint of $H_0=68.6_{-1.2}^{+1.4}~\mbox{km~s}^{-1}~\mbox{Mpc}^{-1}$, $\mytilde50$\% reduction in 1-$\sigma$ errors with respect to the WMAP9-only constraint $H_0=70.0\pm2.2~\mbox{km~s}^{-1}~\mbox{Mpc}^{-1}$. This value is further tightened to $H_0=68.5_{-1.1}^{+0.6}~\mbox{km~s}^{-1}~\mbox{Mpc}^{-1}$ when the BAO result is added.

In case of the non-flat $\Lambda$CDM universe, the WMAP9 result alone cannot determine the $H_0$ value tightly; Hinshaw et al. (2013) quote $38~\mbox{km~s}^{-1}~\mbox{Mpc}^{-1}<H_0<84~\mbox{km~s}^{-1}~\mbox{Mpc}^{-1}$ at the 95\% confidence. This degeneracy is lifted when the DLS cosmic shear is incorporated, giving 
$H_0=68.0_{-5.2}^{+4.0}~\mbox{km~s}^{-1}~\mbox{Mpc}^{-1}$.
We obtain $H_0=68.3_{-1.4}^{+0.6}~\mbox{km~s}^{-1}~\mbox{Mpc}^{-1}$, combining the DLS+BAO+WMAP9 data.

For the $w$CDM cosmology, meaningful constraints are achieved only when we use the DLS+BAO+WMAP9 joint analysis. We measure $H_0=67.8_{-0.9}^{+3.3}~\mbox{km~s}^{-1}~\mbox{Mpc}^{-1}$  and  $70.1_{2.3}^{+4.1}~\mbox{km~s}^{-1}~\mbox{Mpc}^{-1}$ for the flat and curved $w$CDM universes, respectively.

The contribution of the supernova data is most significant for $w$CDM, where the Hubble constant is severely degenerate with other parameters even after we combine the DLS, WMAP9, and BAO data; for $\Lambda$CDM, the degeneracy 
still remains without supernova constraint, but is substantially reduced. Interestingly, the result $H_0=0.697_{-0.004}^{+0.003}$ from the addition of the supernova data corresponds to the 1$\sigma$ upper-limit of the DLS+WMAP9+BAO result.

\subsubsection{Scalar Spectral Index $n_s$ and Baryon Fraction $\Omega_b$}

\begin{figure*}
\centering
\includegraphics[width=8.8cm,trim=0.8cm 0cm 0cm 0cm]{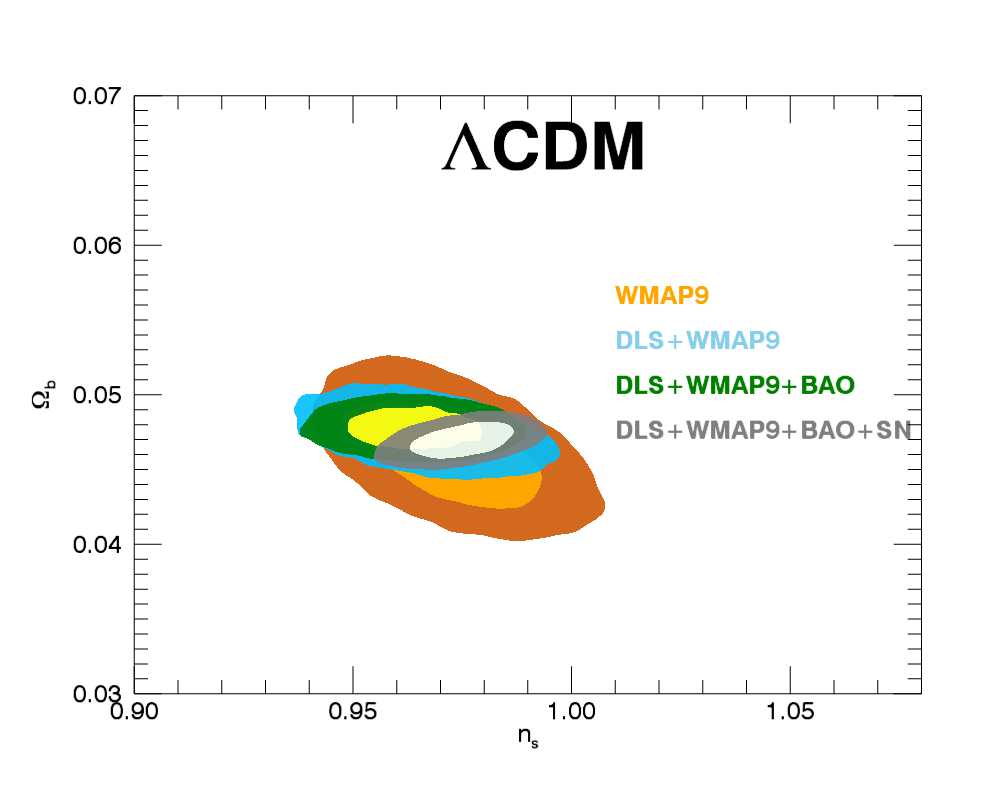}
\includegraphics[width=8.8cm,trim=0.8cm 0cm 0cm 0cm]{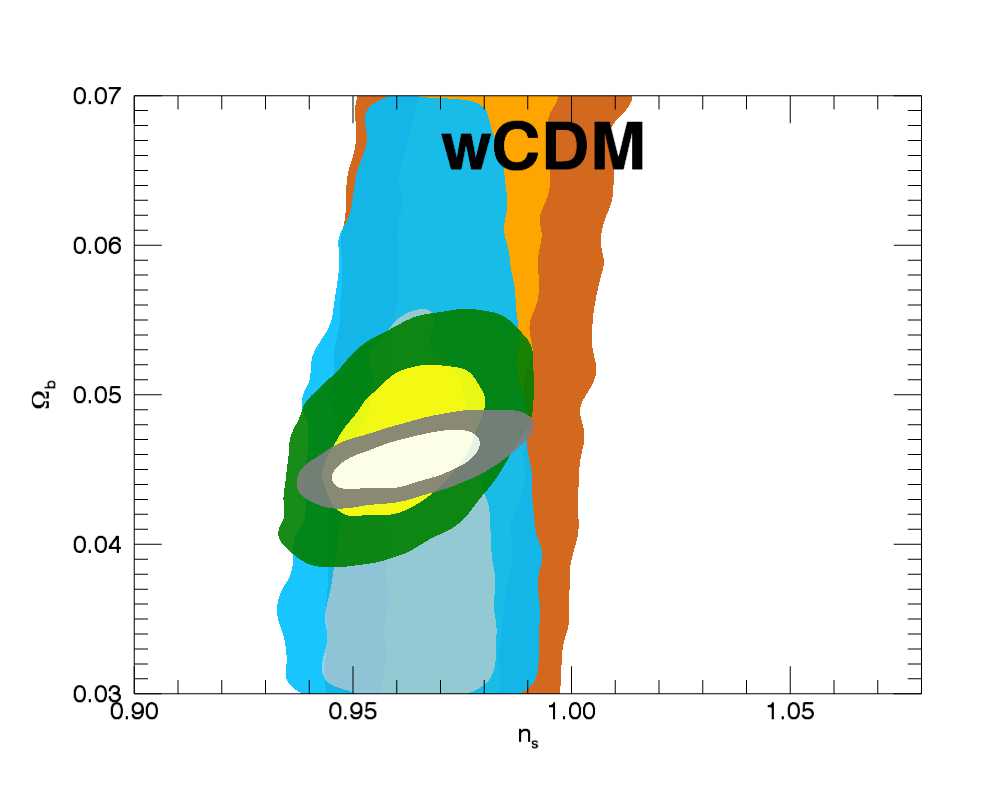}
\includegraphics[width=8.8cm,trim=0.8cm 0cm 0cm 0cm]{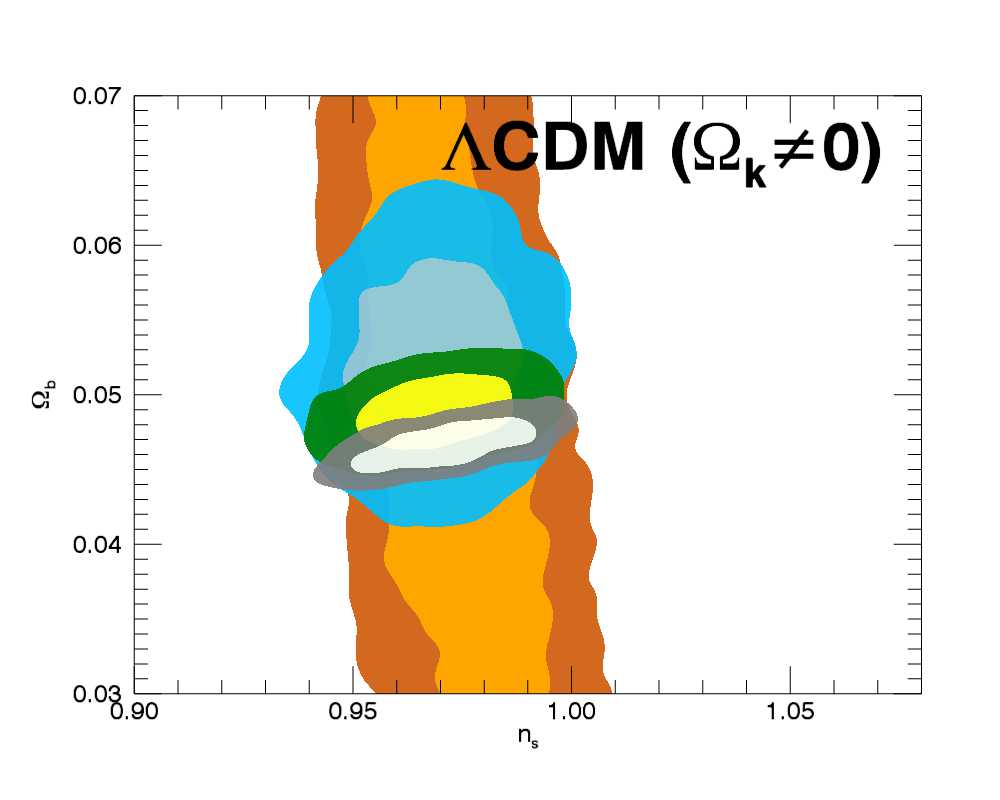}
\includegraphics[width=8.8cm,trim=0.8cm 0cm 0cm 0cm]{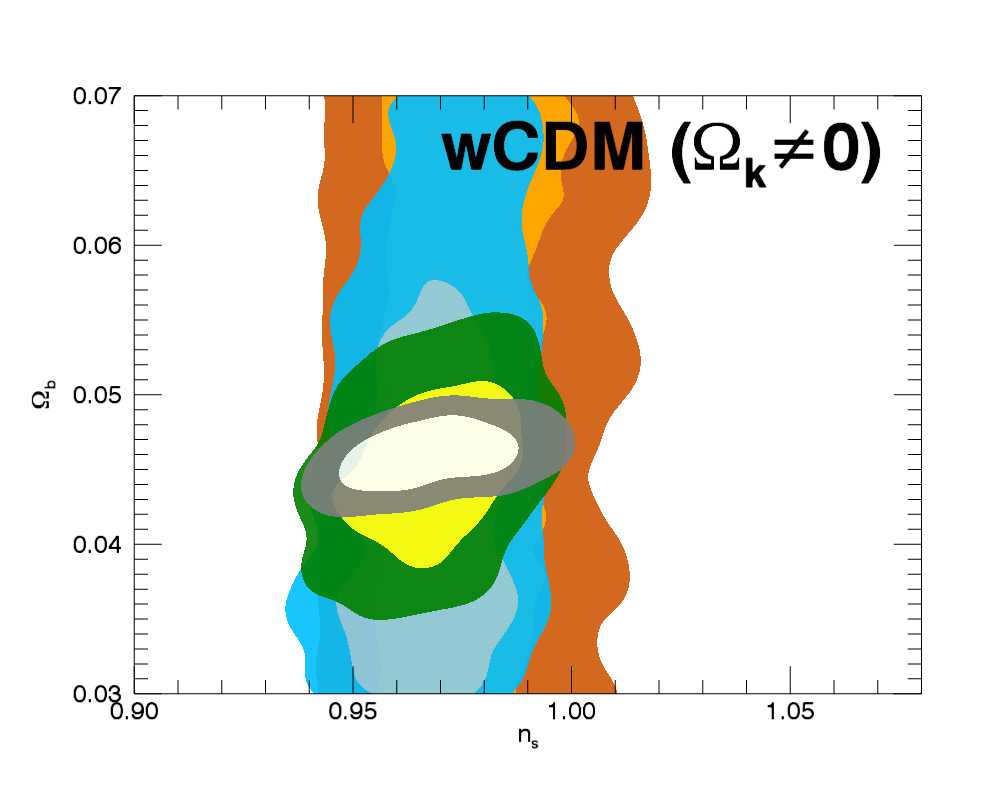}
\caption{Constraint on $n_s$ and $\Omega_b$.}
\label{fig_ns_omega_b}
\end{figure*}

The WMAP9 data alone already put tight constraints on $n_s=0.972\pm0.013$ and $\Omega_b=0.0463\pm0.0024$ for $\Lambda$CDM (Hinshaw et al. 2013). These constraints are further tightened
to $n_s=0.966_{-0.013}^{+0.010}$ and $\Omega_b=0.0475_{-0.0012}^{+0.0013}$ with the addition
of the DLS cosmic shear (Figure~\ref{fig_ns_omega_b}). Supplementing the WMAP9+DLS analysis with the BAO data slightly improves the constraints for this standard flat $\Lambda$CDM cosmology, giving $n_s=0.965_{-0.012}^{+0.008}$ and $\Omega_b=0.0478_{-0.0010}^{+0.0009}$.

For the non-flat $\Lambda$CDM universe, the WMAP9 data can only determine the spectral index $n_s$, and the baryon fraction $\Omega_b$ is not well-constrained. Adding the DLS cosmic shear to the WMAP9 data enables us to constrain the baryon fraction $\Omega_b=0.0518_{-0.0044}^{+0.0049}$ whereas the constraint on the spectral index is hardly affected $n_s=0.968_{-0.013}^{+0.012}$; the WMAP9-only value is $n_s=0.969\pm0.014$.
When the BAO result is included, we can reduce the 1-$\sigma$ uncertainty for $\Omega_b$
by $\mytilde62$\%. However, again the constraint on the spectral index $n_s$ is not improved for this cosmology.

For the $w$CDM cosmology, useful constraints on $\Omega_b$ are obtained only when the three probes (DLS+BAO+WMAP9) are combined at least; we obtain $\Omega_b=0.0469_{-0.0033}^{+0.0036}$ and $0.0437_{-0.0026}^{+0.0056}$ for the flat and the curved $w$CDM cosmologies, respectively. 
The WMAP9 data alone gives $n_s=0.975\pm0.015$, and adding the DLS or the BAO data
does not improve the accuracy of this parameter.

Supernova data alone can only probe geometrical effects and do not constrain the baryonic fraction $\Omega_b$ nor the spectral index $n_s$ directly. However, when combined with other probes, the data can shrink the two parameter uncertainties through their degeneracy with other parameters. The improvement is greater in $\Omega_b$ than in $n_s$
and also for $w$CDM than for $\Lambda$CDM. For example, for a flat $w$CDM, the addition of the supernova data
gives $\Omega_b=0.0455_{-0.0012}^{+0.0014}$, a $\mytilde60$\% reduction in uncertainty compared to the result $\Omega_b=0.0.0469_{-0.0033}^{+0.0036}$ obtained without the supernova data.

\subsubsection{Curvature $\Omega_k$}
\begin{figure*}
\centering
\includegraphics[width=8.8cm,trim=0.8cm 0cm 0cm 0cm]{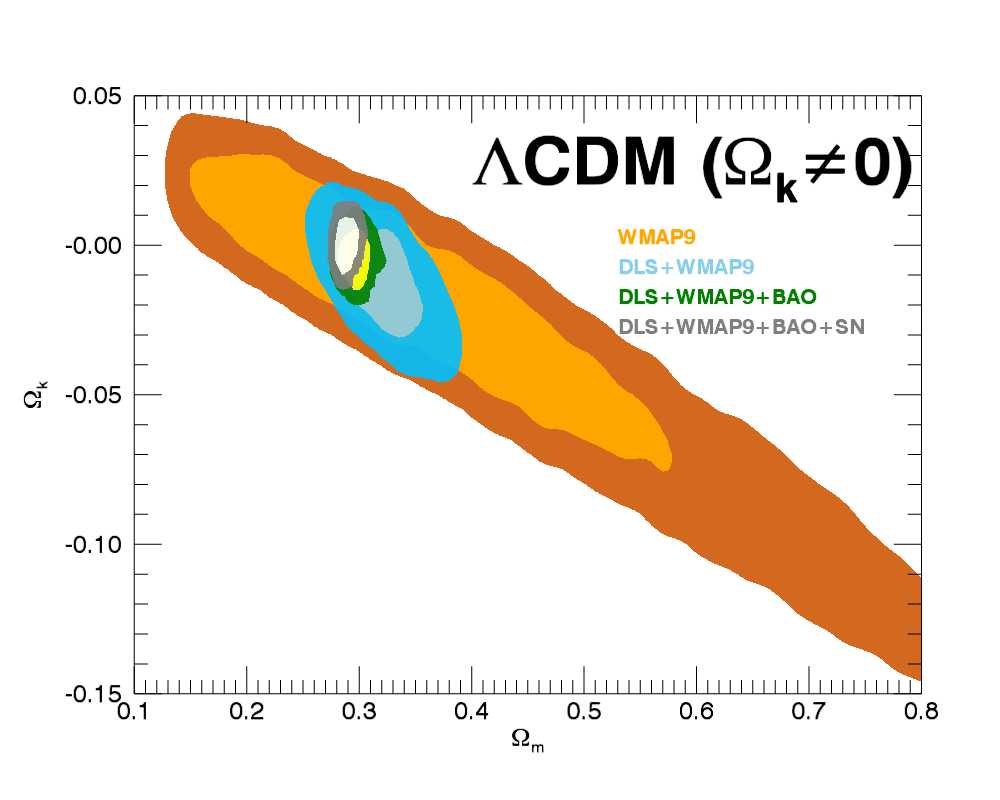}
\includegraphics[width=8.8cm,trim=0.8cm 0cm 0cm 0cm]{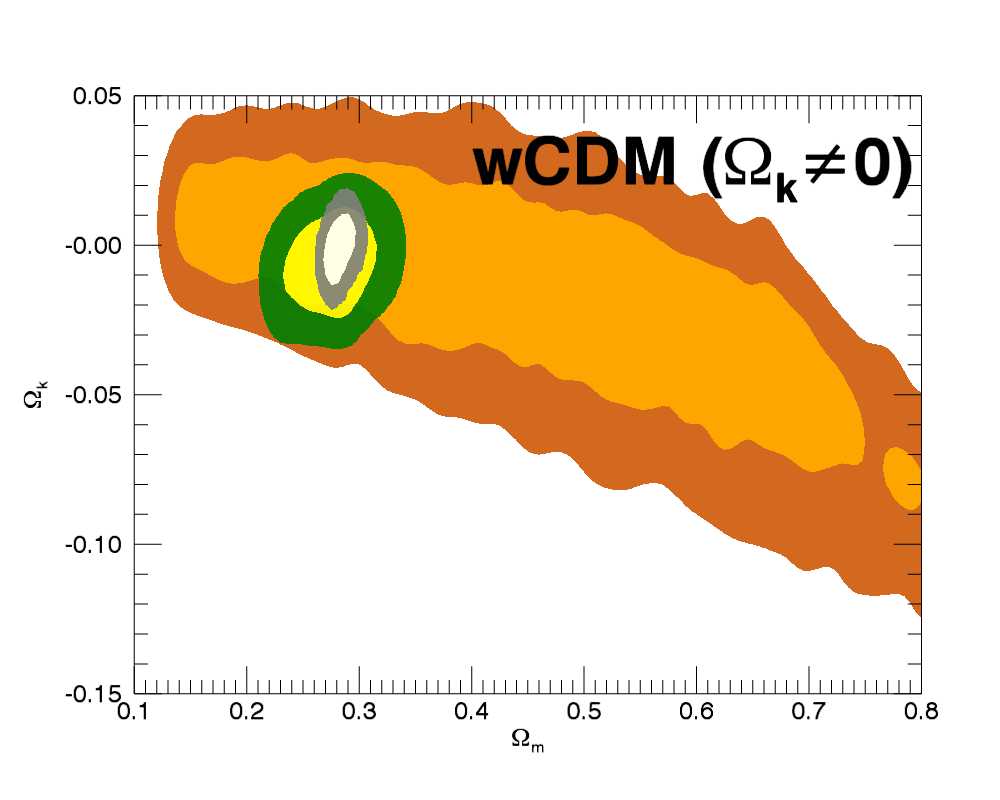}
\caption{Constraint on $\Omega_k$.}
\label{fig_omega_k}
\end{figure*}

The curvature parameter $\Omega_k$ is only loosely constrained by WMAP9 alone for $\Lambda$CDM: $-0.212<\Omega_k<0.021$ at the 95\% confidence (Hinshaw et al. 2013).
When the WMAP9 result is supplemented with the DLS cosmic shear, we achieve
a remarkable improvement $\Omega_k=-0.010_{-0.015}^{+0.013}$ (Figure~\ref{fig_omega_k}), which
is further tightened to $\Omega_k=-0.004_{-0.006}^{+0.005}$ with the addition of the BAO result. 

For $w$CDM, the role of the DLS cosmic shear is insignificant in constraining $\Omega_k$. From the DLS+BAO+WMAP9, we obtain $\Omega_k=-0.006\pm0.011$, which is only a $\mytilde10\%$ reduction in the 1-$\sigma$ uncertainty compared to the WMAP9+BAO result. The contribution of the supernova data is negligible for $\Lambda$CDM and marginal for $w$CDM  when constraining $\Omega_k$.

\subsubsection{Dark Energy Equation of State $w$}

\begin{figure*}
\centering
\includegraphics[width=8.8cm,trim=0.8cm 0cm 0cm 0cm]{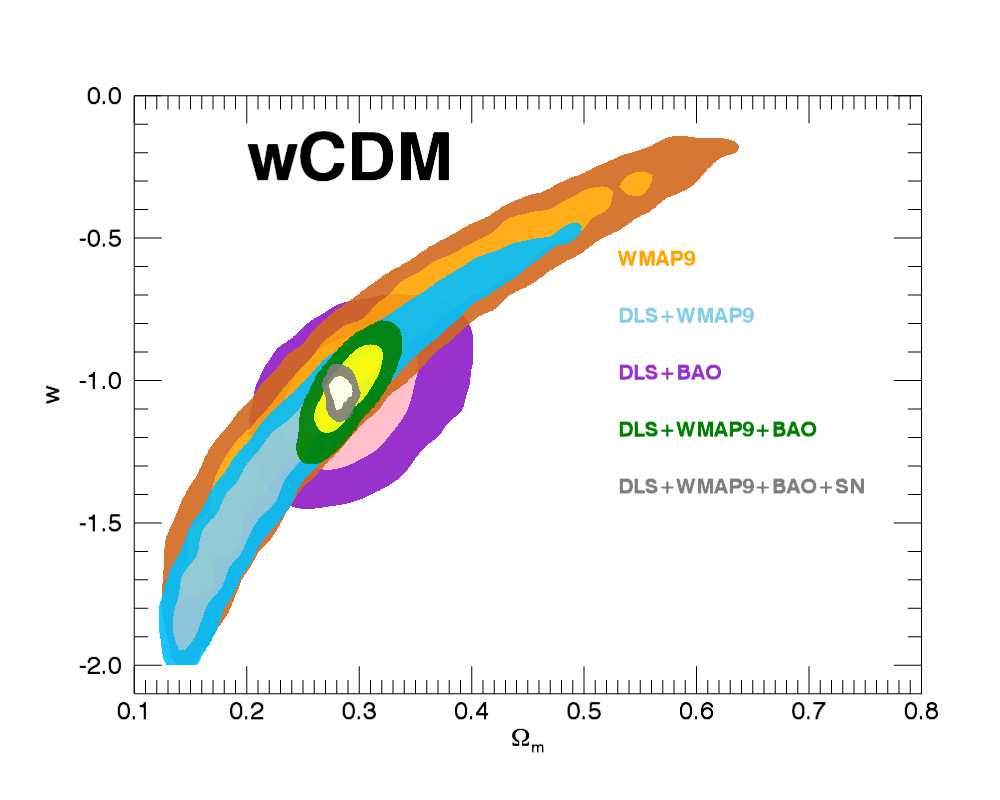}
\includegraphics[width=8.8cm,trim=0.8cm 0cm 0cm 0cm]{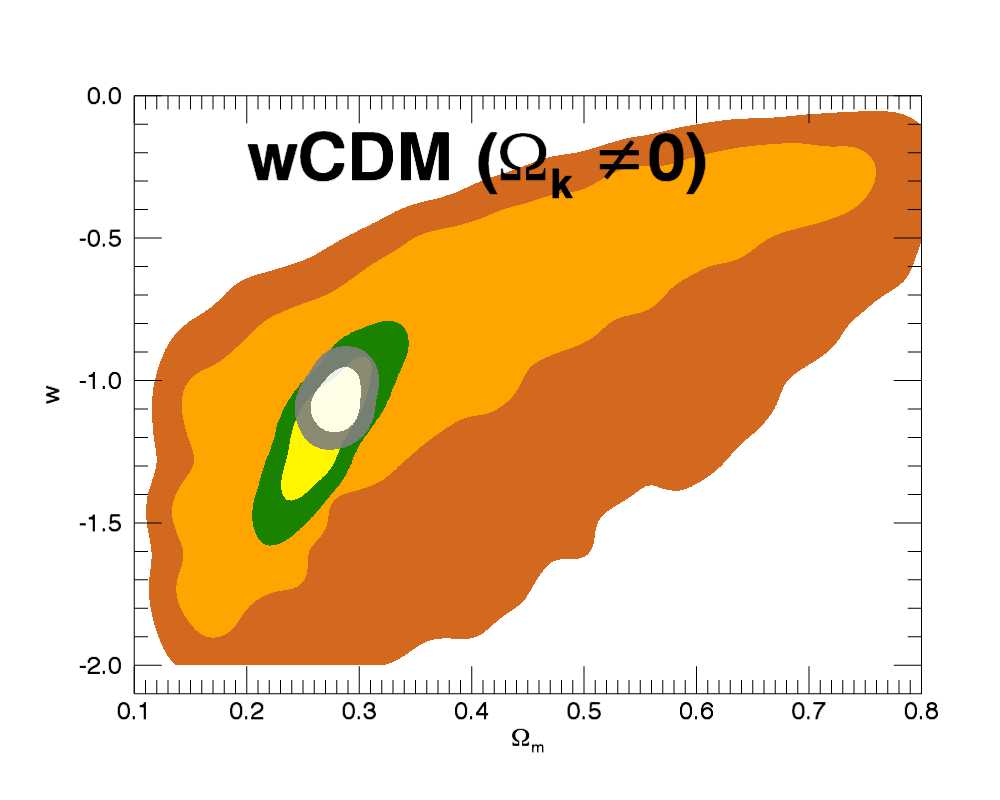}
\caption{Constraint on $w_0$.}
\end{figure*}

The parameter $w$ relates the fluid pressure $p$ to its energy density $\rho$ via $w=p/\rho$ with 
$w=-1$ corresponding to a constant dark energy density or $\Lambda$CDM. The WMAP9 data alone does not constrain $w$ well because the effect of dark energy at the last scattering surface $z\sim1100$ is not significant. The effect is significant in ``low-$z$" probes such as cosmic shear or BAO.  Nevertheless, the DLS data alone cannot constrain the lower-limit of $w$ because of its degeneracy largely with the normalization $\sigma_8$. Needless to say, the BAO data alone cannot determine $w$ either because it provides only a distance-measure at a single redshift.

For the flat $w$CDM, the WMAP9+DLS joint analysis breaks the degeneracy and yields $w=-1.54_{-0.18}^{+0.55}$. Also, the combination of the DLS cosmic shear with the BAO data
yields $w=-1.06_{-0.15}^{+0.17}$. From the joint analysis using the three DLS+WMAP9+BAO probes, we obtain a tight constraint $w=-1.02_{-0.09}^{+0.10}$.

When the curvature constraint is relaxed ($\Omega_k \neq 0$), useful constraints are only obtained when both WMAP9 and BAO data are included. The DLS+WMAP9+BAO joint analysis
gives $w=-1.13_{-0.21}^{+0.13}$, which is $\mytilde15$\% improvement on accuracy
compared to the WMAP9+BAO result $w=-1.05_{-0.19}^{+0.21}$ (Hinshaw et al. 2013).

The addition of the supernova data improves the constraint on $w$ by a factor of 3-4. With and without the curvature constraint $\Omega_k\equiv0$, we obtain $w=-1.03\pm0.03$ and $-1.09_{-0.07}^{+0.09}$, respectively. 

\begin{deluxetable*}{clcccccc}
\tabletypesize{\scriptsize}
\tablecaption{Summary of cosmological parameter constraints from joint probes.}
\tablenum{2}
\tablehead{\colhead{Parameter} & \colhead{Joint probe} &  \colhead{$\Lambda$CDM  $(\Omega_k \equiv 0)$} & \colhead{$\Lambda$CDM $(\Omega_k \neq 0)$}   & \colhead{$w$CDM  $(\Omega_k \equiv 0)$} & \colhead{$w$CDM $(\Omega_k \neq 0)$}    }
\tablewidth{0pt}
\startdata
$\Omega_m$ & DLS+BAO       &  $0.291_{-0.035}^{+0.039}$  & $0.291_{-0.033}^{+0.039}$ &  $0.286_{-0.037}^{+0.043}$ &  $0.259_{-0.047}^{+0.052}$  \\
           & DLS+WMAP9        &  $0.293_{-0.014}^{+0.012}$  & $0.315_{-0.024}^{+0.038}$ & $0.191_{-0.051}^{+0.085}$ &  - \\
           & DLS+BAO+WMAP9 &  $0.297_{-0.012}^{+0.010}$  & $0.297_{-0.012}^{+0.011}$ & $0.290_{-0.017}^{+0.020}$  & $0.269_{-0.024}^{+0.029}$   \\
        & DLS+BAO+WMAP9+SN & $0.283_{-0.005}^{+0.007}$  & $0.286_{-0.011}^{+0.009}$ & $0.286_{-0.011}^{+0.008}$  & $0.279_{-0.009}^{+0.012}$   \\[0.05in]
        
$\sigma_8$ & DLS+BAO       &  $0.827_{-0.058}^{+0.064}$  & $0.827_{-0.068}^{+0.059}$ & $0.831_{-0.061}^{+0.060}$ &  $0.908_{-0.108}^{+0.092}$ \\
           & DLS+WMAP9        &  $0.833_{-0.018}^{+0.011}$  & $0.805_{-0.025}^{+0.025}$ & $0.922_{-0.091}^{+0.129}$ &  - \\
           & DLS+BAO+WMAP9 &  $0.833_{-0.018}^{+0.011}$  & $0.837_{-0.013}^{+0.022}$ & $0.845_{-0.039}^{+0.025}$ &  $0.853_{-0.033}^{+0.051}$ \\
       & DLS+BAO+WMAP9+SN &  $0.837_{-0.015}^{+0.013}$  & $0.841_{-0.016}^{+0.010}$ & $0.841_{-0.011}^{+0.022}$ &  $0.849_{-0.017}^{+0.026}$ \\[0.05in]

$n_s$      & DLS+WMAP9        &  $0.966_{-0.013}^{+0.010}$  & $0.968_{-0.013}^{+0.012}$ & $0.962_{-0.011}^{+0.014}$  & $0.968_{-0.013}^{+0.012}$   \\
           & DLS+BAO+WMAP9 &  $0.965_{-0.012}^{+0.008}$  & $0.967_{-0.010}^{+0.014}$    & $0.961_{-0.012}^{+0.012}$  & $0.967_{-0.011}^{+0.015}$   \\
         & DLS+BAO+WMAP9+SN &  $0.978_{-0.010}^{+0.006}$  & $0.974_{-0.016}^{+0.011}$    & $0.961_{-0.009}^{+0.013}$  & $0.962_{-0.009}^{+0.018}$   \\[0.05in]

$\Omega_b$ & DLS+WMAP9        &  $0.0475_{-0.0012}^{+0.0013}$ & $0.0518_{-0.0044}^{+0.0049}$  & $0.0330_{-0.0030}^{+0.0138}$  & $0.0366_{-0.0063}^{+0.0086}$ \\
           & DLS+BAO+WMAP9 &  $0.0478_{-0.0010}^{+0.0009}$ & $0.0487_{-0.0016}^{+0.0018}$  & $0.0469_{-0.0033}^{+0.0036}$  & $0.0437_{-0.0026}^{+0.0056}$ \\
        & DLS+BAO+WMAP9+SN &  $0.0469_{-0.0007}^{+0.0008}$ & $0.0467_{-0.0012}^{+0.0011}$  & $0.0455_{-0.0012}^{+0.0014}$  & $0.0461_{-0.0017}^{+0.0015}$ \\[0.05in]
         
$h$        & DLS+WMAP9        &  $0.686_{-0.012}^{+0.014}$  & $0.680_{-0.052}^{+0.040}$   & -  & -     \\
           & DLS+BAO+WMAP9 &  $0.685_{-0.011}^{+0.006}$  & $0.683_{-0.014}^{+0.006}$   &  $0.678_{-0.009}^{+0.033}$  &  $0.701_{-0.023}^{+0.041}$  \\
      & DLS+BAO+WMAP9+SN &  $0.697_{-0.004}^{+0.003}$  & $0.697_{-0.004}^{+0.004}$   &  $0.701_{-0.006}^{+0.006}$  &  $0.697_{-0.004}^{+0.009}$  \\[0.05in]

$\Omega_k$ & DLS+WMAP9        &  0                         & $-0.010_{-0.015}^{+0.013}$ & 0 & -  \\            
        & DLS+BAO+WMAP9 &  0                         & $-0.004_{-0.006}^{+0.005}$ & 0 & $-0.006_{-0.011}^{+0.011}$ \\
        & DLS+BAO+WMAP9+SN &  0                         & $-0.001_{-0.005}^{+0.006}$ & 0 & $-0.001_{-0.009}^{+0.009}$ \\[0.05in]

$w$      & DLS+BAO            &  -1 & -1 & $-1.06_{-0.15}^{+0.17}$ & -  \\
           & DLS+WMAP9        &  -1 & -1 & $-1.54_{-0.18}^{+0.55}$ & -  \\
           & DLS+BAO+WMAP9    &  -1 & -1 & $-1.02_{-0.09}^{+0.10}$ & $-1.13_{-0.21}^{+0.13}$  \\
      & DLS+BAO+WMAP9+SN    &  -1 & -1 & $-1.03_{-0.03}^{+0.03}$ & $-1.09_{-0.07}^{+0.09}$  

\enddata
\end{deluxetable*}

\section{DISCUSSION} \label{section_discussion}

\subsection{Impact of Intrinsic Alignment Model \label{section_IA_impact}}
In \textsection\ref{section_IA_model} we described the luminosity-dependent IA correction scheme developed by Joachimi et al. (2011), along with the earlier luminosity-independent correction scheme. Here we discuss how the correction scheme affects our parameter constraint, particularly in the $\Omega_m$-$\sigma_8$ plane.

In the luminosity-independent model, the parameter $A$ in equation~\ref{eqn_IA} determines the global amplitude of the IA signal. We consider the case where $A=1$ because this value is favored in the SuperCOSMOS data (Bridle \& King 2007) and thus often referred to as the fiducial model. The case for $A=2$ is included in order to examine how the increased amplitude in IA affects the parameter constraint.
We also consider the $A=-1$ case because although this is not physical, the comparison with the $A=1$ case is heuristic.   

\begin{figure}
\centering
\includegraphics[width=8.8cm,trim=0.8cm 0cm 0.4cm 0.1cm]{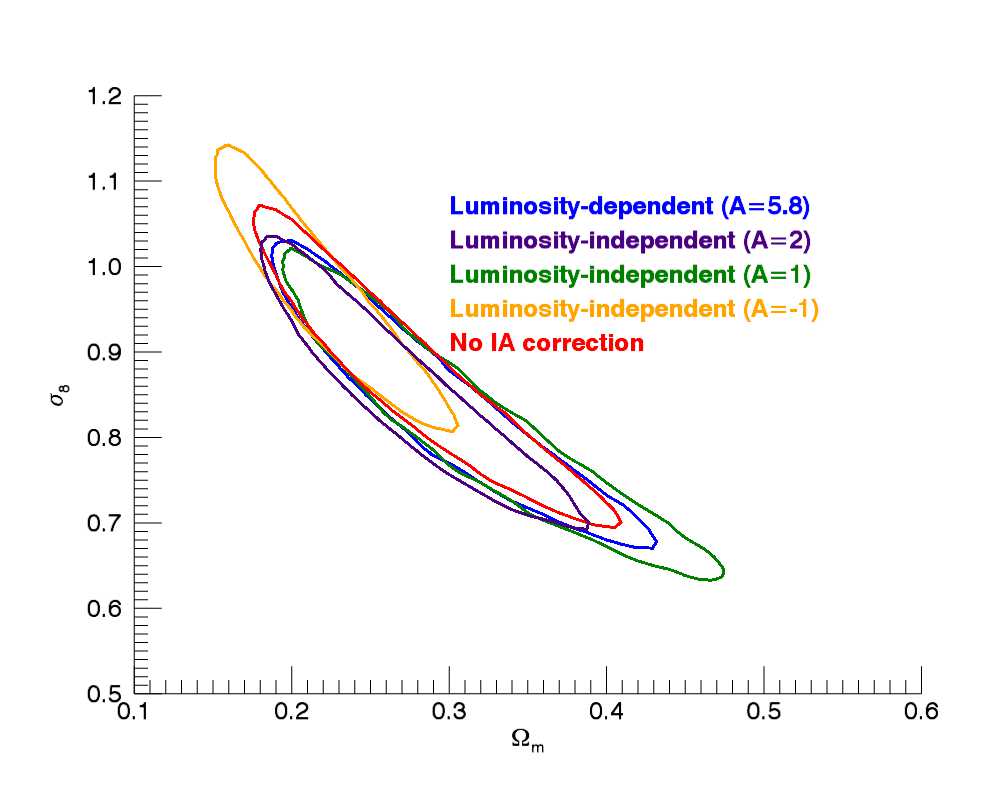}
\caption{Impact of IA correction on DLS cosmic shear. We display the cases for three luminosity-independent IA models with $A=2$ and $A=\pm1$, as well as the luminosity-dependent model (Joachimi et al. 2011). Also illustrated is the result when no IA correction is applied.
We only show 1-$\sigma$ contours to avoid clutter.
In the $\Omega_m-\sigma_8$ plane, the five cases produce consistent results, although we note that different IA schemes move the location of the contours along the $\sigma_8 \Omega_m^a=$constant degeneracy curve.}
\label{fig_impact_of_IA}
\end{figure}

We display the results in Figure~\ref{fig_impact_of_IA}. The comparison shows that 
the five cases produce consistent results, although we note that different IA schemes move the location of the contours along the $\sigma_8 \Omega_m^a=$constant degeneracy curve.
This is somewhat expected because in general IA signals are an order of magnitude smaller than
cosmic shear signals (Figure~\ref{fig_xi_p_all}) and thus do not greatly affect the amplitude of the combination $\sigma_8 \Omega_m^a$. The shifts along the $\sigma_8 \Omega_m^a=$constant degeneracy for different IA schemes are also found in Heymans et al. (2013).

In our study, it is difficult to find a trend in the relation between the IA scheme and the direction of the shift. For example, the $A=1$ case favors a combination of high $\Omega_m$
and low $\sigma_8$ values with respect to the case where we completely ignore IA correction.
If this trend continues, increasing the strength of IA by a factor of two (i.e., $A=2$) may shift the center of the contour further lower right. Our experiment shows that
the direction of the shift is reversed in the $A=2$ case, and the result favors a
combination of low $\Omega_m$ and low $\sigma_8$ with respect to the non-IA case.
When we choose a negative value for $A$, both II and GI contributions become positive,
corresponding to a scenario wherein our cosmic shear signals are overestimated.
The $A=-1$ case shows a preference for a combination of low $\Omega_m$ and high $\sigma_8$ values.

The luminosity-dependent IA correction affects the $\Omega_m$-$\sigma_8$ constraint only slightly with respect to the baseline (non-IA correction) case. This is somewhat expected
because the $\left ( L/L_0  \right )^{\beta} $ term in equation~\ref{eqn_fzl} is small. For example, at $z=1$, $\left ( L/L_0  \right )^{\beta} \sim 0.2$, which makes
the GI power spectrum $P_{GI}$ low despite the high value of $A=5.76$. Nevertheless, it is non-trivial to predict the direction of the shift, as the exact value of the $\left ( L/L_0  \right )^{\beta} $ term depends on the source luminosity (redshift).

One of the important lessons from the current experiment is that our inability to model the IA contributions prefectly will have only negligible impact on our cosmological parameter
constraints when the DLS cosmic shear is combined with other probes.

\subsection{Impact of Cutoff Angle \label{section_IA_impact}}

As mentioned in \textsection\ref{section_measurement}, we present the results using
our cosmic shear measurements at $\theta\gtrsim1.3\arcmin$ to mitigate the baryonic effect. Until a quantitative consensus becomes available, this passive measure is inevitable for DLS. However, for the next-generation surveys such as LSST, where 
interpretation is no longer limited by statistics, we may be able to introduce some nuisance parameters to model and calibrate out the baryonic effect.

In addition to the baryonic effect, signals at small scales are also sensitive
to other systematic effects such as non-linear power spectrum, 
reduced shear, cosmology-dependent covariance, etc. If the combined effect is large, we may be able to detect it through iteration with varying cutoff angles, below which the signals are discarded.

\begin{figure}
\centering
\includegraphics[width=8.2cm,trim=0.2cm 0cm 0.4cm 0.1cm]{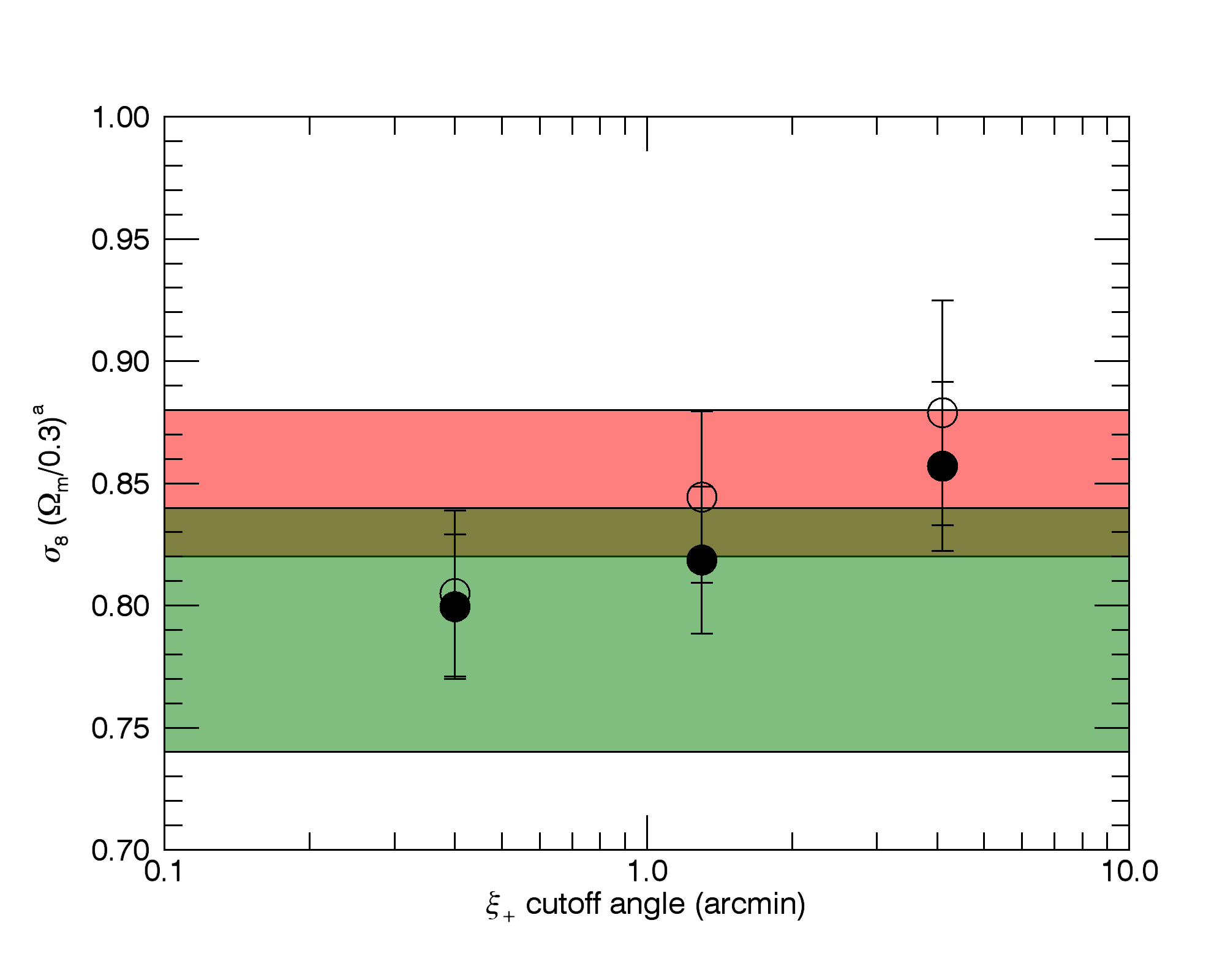}
\caption{Impact of lower-limit cutoff angle. We quantify the difference in results
in terms of the combined normalization $\sigma_8 (\Omega_m/0.3)^a$.
The three filled circles show the results when we use $\theta_{cut}=$0.4$\arcmin$, 1.3$\arcmin$, and 4.1$\arcmin$ for both $\xi_+$ and $\xi_-$. The three open circles represent the results when we applied larger cutoff angles for $\xi_-$; that is, $(\theta^{+}_{cut},\theta^{-}_{cut})=$(0.4$\arcmin$, 4.1$\arcmin$), 
(1.3$\arcmin$, 13.1$\arcmin$), and (4.1$\arcmin$,23.4$\arcmin$).
The reason that one may consider larger cutoff angles for $\xi_-$ is that the $\xi_-$ signals probe smaller structures than the $\xi_+$ signals at the same angular separation $\theta$. If significant, the difference
may hint at small-scale systematics such as baryonic effects, imperfect non-linear power spectrum modeling, etc.
The pink and green boxes represent 1-$\sigma$ ranges bounded by
the Planck2015-CMB and WMAP9 results.
We detect a marginal tendency toward a high normalization with increasing $\theta_{cut}$.
However, given the overlapping error bars, it is difficult to draw any firm conclusions. 
}
\label{fig_cutoff}
\end{figure}

We experiment with different cutoff angles and display the results 
in Figure~\ref{fig_cutoff}. We quantify the difference in results
in terms of the combined normalization $\sigma_8 (\Omega_m/0.3)^a$.
The three filled circles show the results when we use $\theta_{cut}=$0.4$\arcmin$, 1.3$\arcmin$, and 4.1$\arcmin$ for both $\xi_+$ and $\xi_-$. The three open circles represent the results when we applied larger cutoff angles for $\xi_-$; that is, $(\theta^{+}_{cut},\theta^{-}_{cut})=$(0.4$\arcmin$, 4.1$\arcmin$), 
(1.3$\arcmin$, 13.1$\arcmin$), and (4.1$\arcmin$,23.4$\arcmin$).
The reason that one may consider larger cutoff angles for $\xi_-$ is that the $\xi_-$ signals probe smaller structures than the $\xi_+$ signals at the same angular separation $\theta$.

This higher sensitivity of $\xi_-$ to baryonic effects is demonstrated in detail by
the Dark Energy Survey Collaboration et al. (2015), who investigated the impact of the matter power spectrum uncertainty by comparing predicted lensing signals from the OWLS ``AGN" simulations (Schaye et al. 2010) with the dark-matter-only simulation result. 
The OWLS ``AGN" simulation in general predicts lower signal amplitudes than the dark-matter-only simulation.
In particular, the discrepancy in $\xi_-$ is shown to be significant, giving a fractional difference of $\mytilde 10$\% at $\theta\sim20\arcmin$ whereas
the discrepancy in $\xi_+$ is much smaller (even at $\theta\sim2\arcmin$ the fractional difference is only a few percent). 

Having observed the comparison made in the Dark Energy Survey Collaboration et al. (2015), 
one may regard the results as indicating a possibility that the DLS signals at small angles progressively favor a low normalization perhaps due to the baryonic effect.
However, it is important to realize that the six data points in Figure~\ref{fig_cutoff} are still statistically consistent.
We believe that even without any scale-dependent systematics, this kind of a weak trend can still be detected via sample variance (i.e., our DLS fields can happen to contain more large scale structures by chance).

Therefore, for the current study, we can only conclude that the combined systematic effect (e.g., baryons, non-linearity, etc.) at small angles is not large enough to manifest itself beyond the statistical limits. We note that the six results are consistent with both Planck2015-CMB and WMAP9 results.

\subsection{Dependence on the Choice of Nonlinear Power Spectrum Model}
In this paper, we use the Smith et al. (2003) halofit model in our computation of the nonlinear power spectrum. This halofit approach approximates the ``real" power spectrum sampled from $N$-body simulation data and has been widely applied to a number of cosmology studies. However, recent studies report that the model needs to be updated to meet the requirements of future cosmology surveys (e.g., Heitmann et al. 2010; Kiessling et al. 2011). One obvious weakness of the Smith et al. (2003) model arises from the resolution of the numerical simulation that the authors employed. Although considered state-of-the-art at the time of the publication,  it is possible to easily improve the fidelity of the simulation considerably these days. The consensus regarding the impact of the insufficient resolution of the Smith et al. (2002) simulation is that the model consistently underpredicts the power on small scales (e.g., Hilbert et al. 2009; Casarini et al. 2012; Sato et al. 2009; Boylan-kolchin et al. 2012).

Recently, Takahashi et al. (2012) updated the Smith et al. (2003) halofit model coefficients based on their higher-resolution $N$-body simulations; the total number of parameters was increased from 30 (the original) to 35 to increase the degree of freedom in the model and thus to
improve the agreement. Their result supports the claim that the old Smith et al. (2003) model prediction is lower than the high-resolution result non-negligibly for future cosmic shear surveys such as DES and LSST. For instance, at $z=0.35$ the old version underestimates the power at $k=1~h~\mbox{Mpc}^{-1}$ by $\mytilde5$\%. The direction of the shift implies that cosmic shear studies with the old Smith et al. (2003) halofit model lead to overestimation of the normalization.
Indeed, MacCrann et al. (2015) showed that their re-analysis of the CFHTLenS data with the Takahashi et al. (2012) model lowers $\sigma_8$ by $\mytilde0.02$. Our DLS study also supports the findings of MacCrann et al. (2015), decreasing the $\sigma_8$ value at $\Omega_M\equiv0.3$ by $\mytilde0.02$. The shift is not totally negligible in our DLS study, where the parameter constraint even with DLS alone is tight [$\sigma_8 (\Omega_m/0.3)^a=0.818_{-0.026}^{+0.034}$]. However, we defer further analysis until we obtain better understanding of the various barynoic effects on small scales, whose direction and amplitude are currently unknown.

\subsection{Comparison with the Planck2015-CMB Results} \label{section_comparison_with_planck2015}
The Planck2015-CMB paper presents the full-mission results with improved precision over the previous versions. With the Planck data alone, cosmological parameters are tightly constrained, favoring a flat $\Lambda$CDM universe. Because our joint probe using the DLS, WMAP9, and BAO data provide independent constraints with similar precision, here we provide somewhat detailed comparisons.
The Planck2015-CMB paper list results from different combinations of probes. We quote the results obtained from the combination: temperature (TT) + polarization (lowP) + lensing. 

As already discussed in \textsection\ref{section_DLS_only}, the Planck2015 constraint on the combination $\sigma_8 (\Omega_m/0.3)^a=0.825\pm0.016$ is in good agreement with our DLS-only result $\sigma_8 (\Omega_m/0.3)^a=0.818_{-0.026}^{+0.034}$ (current tomography) and $\sigma_8 (\Omega_m/0.3)^a=0.82\pm0.04$ (DLS 2D; Jee et al. 2013).
This Planck result is however at tension with the CFHTLenS-only measurement $\sigma_8 (\Omega_m/0.3)^a=0.74_{-0.04}^{+0.03}$ (Heymans et al. 2013), as also noted by MacCrann et al. (2014). 

The matter density and normalization values from Planck2015 are $\Omega_m=0.308\pm0.012$ and $\sigma_8=0.815\pm0.009$, in good 
agreement with our joint probe results $\Omega_m=0.297_{-0.012}^{+0.010}$ and $\sigma_8=0.833_{-0.018}^{+0.010}$.
The scalar spectral index from Planck2015 $n_s=0.9677\pm0.0060$ is also consistent with our measurement $n_s=0.965_{-0.012}^{+0.008}$. A similar level of consistency is found for the baryon fraction $\Omega_b=0.0478\pm0.0009$ (DLS+BAO+WMAP9) versus $\Omega_b=0.0484 \pm 0.0005$ (Planck2015).

The comparison of the Hubble constant $H_0$ is potentially interesting because the Planck measurement $H_0=67.8\pm0.9~\mbox{km~s}^{-1}~\mbox{Mpc}^{-1}$ is
at slight tension ($\mytilde2.4~\sigma$) with the conventional measurement $H_0 = 73.8\pm2.4 ~\mbox{km~s}^{-1}~\mbox{Mpc}^{-1}$ constrained by HST Cepheid+SNe data (Riess et al. 2011). It is worth mentioning that the $H_0$ constraint with the CMB is indirect despite a smaller statistical uncertainty. If the tension persists, the discrepancy may indicate a non-negligible incompleteness in our model regarding the CMB mechanism and/or the structure evolution.
Although cosmic shear (like other low-redshift probes) is sensitive to the $Hubble$ constant because it is used for the translation of observed angular scales to physical scales, in general cosmic shear alone cannot bracket the value because of its degeneracy with other parameters. As discussed in \textsection\ref{section_hubble}, the combination of the DLS tomography with
WMAP9 reduces the 1$\sigma$ uncertainty of the WMAP9-alone measurement by $\mytilde50$\%, giving 
$H_0=68.6_{-1.2}^{+1.4} ~\mbox{km~s}^{-1}~\mbox{Mpc}^{-1}$. The central value is about 1-$\sigma$ lower than the WMAP9-alone result and in a better agreement with the Planck2015-CMB measurement. The DLS+BAO+WMAP9 analysis still favors the Planck2015-CMB value,
further tightening the constraint $H_0=68.5_{-1.1}^{+0.6} ~\mbox{km~s}^{-1}~\mbox{Mpc}^{-1}$. 
We refer readers to Planck2015-CMB for further discussions on discrepant $H_0$ measurements.

The curvature measurement $\Omega_k=-0.004_{-0.006}^{+0.005}$ with the DLS+BAO+WMAP9 joint analysis agrees nicely with that from Planck2015 $\Omega_k=-0.005_{-0.017}^{+0.016}$ with the uncertainty of the DLS+BAO+WMAP9 measurement
a factor of three smaller.

As mentioned earlier, the CMB data alone do not provide useful constraints on the dark energy equation of state parameter $w=-1.54_{-0.50}^{+1.62}$. Planck2015 present a few $w$ measurements obtained in conjunction with external data. Combining the TT+lowP+lensing data with BAO, $H_0$, and the Joint Lightcurve Analysis (JLA) measurement gives $w = −1.006_{-0.091}^{+0.085}$ (95\% confidence limits), which is in excellent agreement with the DLS+BAO+WMAP9 result $w=-1.02_{-0.09}^{+0.10}$ (95\% confidence limits).

\section{SUMMARY AND CONCLUSIONS} \label{section_conclusion}
We have presented full cosmological parameter constraints from the DLS, which is the deepest ($r_{\mathrm{lim}}\sim27$ and $\left < z \right >\sim1.2$) cosmic shear survey to date among the existing $>10$ sq. deg. surveys. We construct five tomographic bins where bin membership is based on single point estimates and the exact redshift distribution in each bin is estimated by stacking the redshift probability of individual galaxies within each bin. A luminosity-dependent nonlinear model is employed to address intrinsic alignments of source galaxies, and we find that, according to this model, the intrinsic alignment contamination to our cosmic shear measurement is very small, shifting the constraint on the $\Omega_m$-$\sigma_8$ combination by $\mytilde0.2~\sigma$ along the track defined by the $\sigma_8\Omega_m^a=\mbox{constant}$ degeneracy.

When the DLS tomography is used alone, we obtain the constraint $\sigma_8 \left (\Omega_m/0.3 \right )^{0.50\pm0.13}=0.818\pm{-0.026}^{+0.034}$, which is among the tightest constraints from published cosmic shear results. This normalization value from our DLS tomography is in excellent agreement with our previous 2D result. In addition, the measurement is highly consistent with the final results of both WMAP and Planck missions.

For our joint parameter constraints, the latest CMB, BAO, and SNIa results are combined. The DLS+WMAP9 joint probe gives  $\Omega_m=0.293_{-0.014}^{+0.012}$, $\sigma_8=0.833_{-0.018}^{+0.011}$, $H_0=68.6_{-1.2}^{+1.4}~\mbox{km~s}^{-1}~\mbox{Mpc}^{-1}$, and
$\Omega_b=0.0475\pm0.0012$ for $\Lambda$CDM, reducing the uncertainties of the WMAP9-only constraints by $\mytilde50$\%.

When we do not assume flatness for $\Lambda$CDM, we obtain the curvature constraint $\Omega_k=-0.010_{-0.015}^{+0.013}$ from the DLS+WMAP9 combination, which however is not well-constrained when WMAP9 is used alone.
The dark energy equation of state parameter $w$ is tightly constrained when Baryonic Acoustic Oscillation (BAO) data are added, yielding $w=-1.02_{-0.09}^{+0.10}$ with the DLS+WMAP9+BAO joint probe. The addition of supernova constraints further tightens the parameter $w=-1.03\pm0.03$.
Our joint constraints are fully consistent with the final Planck results and also the predictions of a $\Lambda$CDM universe.

In future studies, the current DLS analysis can be improved in several aspects. One of the most important issues is the baryonic effect. Currently, different numerical recipes for simulating the effect provide conflicting results in both amplitude and direction. Some extreme cases show that the departure from the dark matter only power spectrum can be as high as $\mytilde20$\% at $l\sim2000$. If this is indeed the case, we can no longer take into account the baryonic effect using the current passive method (i.e., discarding data on small scales). 
Another aspect of the analysis that needs further investigation is the effect of the cosmology-dependent covariance. Our previous 2D analysis shows that without the implementation, the constraining power is  compromised. Although we choose not to implement the scheme in the current tomographic analysis because of technical difficulties, it will be a useful endeavor to examine the effect even with a small set of covariance matrice spanning a limited range of cosmological parameter values.

\acknowledgments
{M. James Jee acknowledges support for the current research from  the National Research Foundation of
Korea to  the Center for Galaxy Evolution Research and from the Yonsei University New Faculty Research Infrastructure Support Program. 
We thank Perry Gee for carefully managing the DLS database. The development of the StackFit algorithm and its application to the DLS was funded in part by NSF grant AST-1108893 and DOE grants DE-FG02-07ER41505, DE-FG02-91ER40674. The DLS and our systematics reduction R\&D have received major funding from Lucent Technologies Bell Laboratories and from the NSF (grants AST-0134753, AST-0441072, AST-1108893, and AST-0708433). Stefan Hilbert acknowledges support by the NSF grant number AST-0807458-002, and support by the DFG cluster of excellence \lq{}Origin and Structure of the Universe\rq{} (www.universe-cluster.de). Part of this work performed under the auspices of the U.S. Department of Energy by Lawrence Livermore National Laboratory under Contract DE-AC52-07NA27344. This work is based on observations at Kitt Peak National Observatory and Cerro Tololo Inter-American Observatory, which are operated by the Association of Universities for Research in Astronomy (AURA) under a cooperative agreement with the National Science Foundation.}

\end{document}